\documentclass[aps, unsortedaddress, showpacs, twocolumn, amsmath, amssymb, flushbottom]{revtex4-1}

\usepackage{bm}
\usepackage{latexsym}
\usepackage[utf8]{inputenc}
\usepackage[T1]{fontenc}
\usepackage{graphicx}
\usepackage{multirow}
\usepackage[english]{babel}
\usepackage{lmodern}

\newcommand{\f}{\frac}
\newcommand{\h}{\hat}
\newcommand{\be}{\begin{equation}}
\newcommand{\ee}{\end{equation}}
\newcommand{\ba}{\begin{array}}
\newcommand{\ea}{\end{array}}
\newcommand{\bc}{\begin{center}}
\newcommand{\ec}{\end{center}}

\begin{document}

\title{Shear stress in lattice Boltzmann simulations}

\date{\today}

\author{Timm~Kr\"uger}
\email{t.krueger@mpie.de}
\affiliation{Max-Planck-Institut f\"ur Eisenforschung \\ Max-Planck-Str.\ 1, 40237 D\"usseldorf, Germany}
\author{Fathollah~Varnik}
\affiliation{Interdisciplinary Centre for Advanced Materials Simulation \\ Stiepeler Str.\ 129, 44780 Bochum, Germany}
\affiliation{Max-Planck-Institut f\"ur Eisenforschung \\ Max-Planck-Str.\ 1, 40237 D\"usseldorf, Germany}
\author{Dierk~Raabe}
\affiliation{Max-Planck-Institut f\"ur Eisenforschung \\ Max-Planck-Str.\ 1, 40237 D\"usseldorf, Germany}


\begin{abstract}
A thorough study of shear stress within the lattice Boltzmann method is provided. Via standard multiscale Chapman-Enskog expansion we investigate the dependence of the error in shear stress on grid resolution showing that the shear stress obtained by the lattice Boltzmann method is second order accurate. This convergence, however, is usually spoiled by the boundary conditions. It is also investigated which value of the relaxation parameter minimizes the error. Furthermore, for simulations using velocity boundary conditions, an artificial mass increase is often observed. This is a consequence of the compressibility of the lattice Boltzmann fluid. We investigate this issue and derive an analytic expression for the time-dependence of the fluid density in terms of the Reynolds number, Mach number and a geometric factor for the case of a Poiseuille flow through a rectangular channel in three dimensions. Comparison of the analytic expression with results of lattice Boltzmann simulations shows excellent agreement.
\end{abstract}

\maketitle


\section{Introduction}
\label{sec:introduction}

The simulation of fluids using the lattice Boltzmann method (LBM) has become a very important research field in the last two decades \cite{mcnamara_use_1988, higuera_boltzmann_1989, qian_lattice_1992, benzi_lattice_1992} due to its simple and parallelizable implementation and various fields of application \cite{ladd2001lbs, chen_lattice_1998, kandhai_lattice-boltzmann_1998, succi_lattice_2001, sukop_lattice_2005, varnik_wetting_2008, varnik_roughness-induced_2007, varnik_chaotic_2007, varnik_scaling_2006}.

In the literature, the velocity field $\bm u$ of the fluid usually is the central observable of interest. More recently, however, some researchers are becoming interested in the shear stress field $\bm \sigma \propto \bm \nabla \bm u$, in particular when studying blood flow \cite{boyd_application_2005, xiu-ying_three-dimensional_2008}. High shear stresses may promote blood clotting, an effect with quite diverse consequences such as stopping blood flow through injuries (a desired effect) or blocking vessels plagued by arteriosclerosis \cite{schneider_shear-induced_2007}. In the case of diffusive scaling, the time step and the lattice constant obey $\Delta t \propto \Delta x^2$, and the solution to the lattice Boltzmann (LB) equation converges to the solution to the Navier-Stokes equations with a second order rate, i.e., the relative error $e_u$ of the velocity scales with $\Delta x^2$. Noting that, within diffusive scaling, the Mach number linearly scales with grid resolution, $\text{Ma} \propto \Delta x$, one also obtains $e_u \propto \text{Ma}^2$.

To our knowledge, a systematic analysis of the convergence behavior of the shear stress in the LBM has not been conducted yet. It is often stated that the shear stress enters the LB equation on the $\epsilon^2$ level leading to a convergence $\propto \Delta x$ only. As will be shown in this report, this a priori statement underestimates the accuracy of the LB simulations.

Velocity boundary conditions are very useful in the LBM. They allow to impose a desired velocity field at the borders of the computational domains. This is crucial, since the solution to the Navier-Stokes equations critically depends on the boundary conditions. One drawback of those boundary conditions is the unphysical mass increase, observed under some circumstances. In this paper, we study this aspect via analytic means. In particular, using the compressibility of the LB fluid, we solve the Stokes equation for a rectangular channel subject to a Poiseuille flow and show that the fluid density exponentially grows with time. The rate of mass increase is found to depend on the inverse Reynolds number but scales as the third power of the Mach number. We demonstrate the validity of analytic results through comparison with LB simulations.

This article consists of two parts. First, we investigate the convergence of the shear stress in a laminar 3D Poiseuille flow. Moreover we address the issue of optimum parameter choice for the LB simulations. We especially search for a reasonable value for the relaxation parameter $\tau$ in view of the shear stress as the relevant observable. The second part deals with the global mass increase in the simulations, which arises when velocity boundary conditions for both the inlet and outlet of the flow are used. This is particularly important for the correct computation of the shear stress.

The paper is organized as follows: In section \ref{sec:lbm}, the theoretical background is provided. After a short introduction to the LBM in subsections \ref{subsec:basics} through \ref{subsec:shearstress}, the convergence behavior is discussed in subsection \ref{subsec:convergence}. Velocity boundary conditions are briefly dealt with in subsection \ref{subsec:geometry}, and the influence of the initial simulation parameters is addressed in subsection \ref{subsec:relparameter}. The mass increase is analyzed in section \ref{sec:massincrease}. The numerical results are presented in section \ref{sec:simulations}, followed by the conclusions in section \ref{sec:conclusions}.


\section{The lattice Boltzmann method}
\label{sec:lbm}

\subsection{The basic concepts}
\label{subsec:basics}

The LBM has become a very efficient Navier-Stokes solver. Its strength is based on its simple coding and its locality, which makes it intrinsically parallelizable. Only next-neighbor information is required during the evaluation of the LB equation.

The LBM can either be regarded as the discretized form of the Boltzmann equation or as the extension of lattice gas automata \cite{frisch_lattice-gas_1986}. There is one major difference between the LBM and other Navier-Stokes solvers: While conventional methods directly solve the Navier-Stokes equations in terms of the density $\rho$ and the velocity $\bm u$ and approximate the differential equations by finite differences, the LBM introduces a number of $q$ populations $f_i$ ($i = 0, \ldots, q - 1$) streaming along a regular lattice in discrete time steps. Those populations can be regarded as mesoscopic particle packets propagating and colliding. For the entire LBM, only algebraic manipulations of the populations are required. A more detailed presentation of the LBM can be found in the literature (e.g., \cite{succi_lattice_2001, sukop_lattice_2005} and the references therein).

It is natural to consider all following equations in this article as being dimensionless. The physical dimensions are recovered by multiplying the lattice quantities by appropriate combinations of the lattice constant $\Delta x$, the time step $\Delta t$ and the density $\rho$ given in $\text m$, $\text s$, and $\text{kg} / \text{m}^3$, respectively. In lattice units, the time step and lattice constant are usually set to unity.

The evolution of the populations $f_i$ is given by the LB equation, which takes the form
\be
\label{eq:lbequation}
f_i (\bm x + \bm c_i, t + 1) - f_i (\bm x, t) = - \f{1}{\tau} \left(f_i (\bm x, t) - f_i^{\text{eq}} (\bm x, t)\right)
\ee
in the Bhatnagar-Gross-Krook (BGK) approximation. The equilibrium $f_i^{\text{eq}}$ is defined below. $\tau$ is the relaxation time of the LB fluid. At each time step, the populations propagate along the $q$ discretized velocity vectors $\bm c_i$ to the next neighbors. At those points they collide according to the right-hand-side of Eq.\ (\ref{eq:lbequation}). In this article, we use a 3D model with $15$ velocities, designated $D3Q15$. This lattice and the corresponding velocity vectors $\bm c_i$ are introduced in \cite{qian_lattice_1992}. The dimensionless relaxation parameter $\tau$ is connected to the speed of sound $c_s$ and the kinematic viscosity $\nu$ of the fluid by $\nu = c_s^2 (\tau - 1/2)$. In lattice units the sound speed takes the value $c_s^2 = 1 / 3$. In the following we will use Latin indices for the populations $f_i$ and Greek indices for spatial components, e.g., $u_\alpha$. Throughout the article, the Einstein sum convention is used.

The $q$ populations $f_i$ carry the entire information of the fluid, and the macroscopic observables like density and velocity can directly be recovered by
\be
\label{eq:lbmmomenta}
\rho = \sum_{i=0}^{q-1} f_i \,, \quad \rho\, \bm u = \sum_{i=0}^{q-1} \bm c_i f_i
\ee
at every fluid lattice node. Also the deviatoric shear stress $\bm \sigma$ can be computed from the populations. We will come back to this issue in more detail in subsection \ref{subsec:shearstress}. In the following we drop the summation limits.

In Eq.\ (\ref{eq:lbequation}) the equilibrium populations are given by
\be
\label{eq:equilibrium}
f_i^{\text{eq}} = w_i \rho \left(1 + 3 \bm c_i \cdot \bm u + \f 92 (\bm c_i \cdot \bm u)^2 - \f 32 \bm u \cdot \bm u\right) \,.
\ee
This is closely related to the truncated form of the Maxwell distribution, which is a very good approximation for small Mach numbers. Thus, the LBM is only reasonable for small velocities compared to sound speed. The $q$ factors $w_i$ are the lattice weights, depending on the underlying lattice structure. Their choice ensures the isotropy of the fluid, a necessity to solve the Navier-Stokes equations asymptotically. For the $D3Q15$ lattice, they are also given in \cite{qian_lattice_1992}.

Beside the constraint of small velocities, the LBM suffers from another shortcoming: the incompressibility of the fluid gives way to a quasi incompressibility indicated by the ideal gas equation of state, $p = c_s^2 \rho$. As will be exposed in section \ref{sec:massincrease}, this is the reason for the mass increase when using velocity boundary conditions.

In order to show that the LB equation is equivalent to the incompressible Navier-Stokes equations in the limit $\Delta x \to 0$ and $\text{Ma} \to 0$, one performs a Chapman-Enskog analysis, which will be briefly presented in the following subsection.


\subsection{Chapman-Enskog analysis}
\label{subsec:chapmanenskog}

The idea behind the Chapman-Enskog analysis applied to the LB equation is that different physical phenomena happen on different time scales. While the advection of the fluid is the fastest process, the diffusion of mass, momentum and energy happens on a slower time scale. For this reason, the time derivative is usually split into two parts. One can also extend this approach and take into account more time scales,
\be
\label{eq:ce_timescales}
\partial_t = \epsilon \partial_{t_0} + \epsilon^2 \partial_{t_1} + \epsilon^3 \partial_{t_2} + {\cal O} (\epsilon^4) \,.
\ee
Since the spatial variations of all processes are of the same order, the gradient is not decomposed, $\bm \nabla = \epsilon \bm \nabla_1$. The populations are expanded about the equilibrium,
\be
f_i = f_i^{(0)} + \epsilon f_i^{(1)} + \epsilon^2 f_i^{(2)} + \epsilon^3 f_i^{(3)} + {\cal O} (\epsilon^4) \,,
\ee
where $f_i^{(0)} = f_i^{\text{eq}}$. The parameter $\epsilon \ll 1$ can be identified as the Knudsen number (the ratio of mean free path to a characteristic length, usually the size of an obstacle or the entire system). The Taylor expanded LB equation [Eq.\ (\ref{eq:lbequation})] takes the form
\be
\label{eq:ce_lbe}
\sum_{n=1}^\infty \f{1}{n!} \left(\partial_t + \bm c_i \cdot \bm \nabla\right)^n f_i (\bm x, t) = - \f{1}{\tau} \left(f_i (\bm x, t) - f_i^{\text{eq}} (\bm x, t)\right) \,.
\ee
The expansions for $\partial_t$, $\bm \nabla$ and $f_i$ are inserted into Eq.\ (\ref{eq:ce_lbe}), and the terms are sorted by the powers of $\epsilon$. The $\epsilon$-equation reads
\be
\label{eq:ce_firstorder}
(\partial_{t_0} + \bm c_i \cdot \bm \nabla) f_i^{(0)} = - \f{1}{\tau} f_i^{(1)} \,,
\ee
the $\epsilon^2$-equation
\be
\label{eq:ce_secondorder}
\begin{aligned}
\partial_{t_1} f_i^{(0)} + (\partial_{t_0} + \bm c_i \cdot \bm \nabla) f_i^{(1)} \\
+ \f 12 \left(\partial_{t_0} + \bm c_i \cdot \bm \nabla\right)^2 f_i^{(0)} &= - \f{1}{\tau} f_i^{(2)} \,,
\end{aligned}
\ee
and the $\epsilon^3$-equation
\begin{widetext}
\be
\begin{aligned}
\partial_{t_2} f_i^{(0)} + \partial_{t_1} f_i^{(1)} + (\partial_{t_0} + \bm c_i \cdot \bm \nabla) f_i^{(2)}
+ \f 12 \left(\partial_{t_0} + \bm c_i \cdot \bm \nabla\right)^2 f_i^{(1)} \\
+ \f 12 (\partial_{t_0} \partial_{t_1} + \partial_{t_1} \partial_{t_0} + 2 \bm c_i \cdot \bm \nabla \partial_{t_1}) f_i^{(0)}
+ \f 16 \left(\partial_{t_0} + \bm c_i \cdot \bm \nabla\right)^3 f_i^{(0)} &= - \f{1}{\tau} f_i^{(3)} \,.
\end{aligned}
\ee
\end{widetext}

For recovering the Navier-Stokes equations in terms of $\rho$ and $\bm u$, the first and second order equations (\ref{eq:ce_firstorder}) and (\ref{eq:ce_secondorder}) and the same equations multiplied by $\bm c_i$ are summed over $i$. One can show that this approach yields the four macroscopic equations
\be
\label{eq:ce_firstorderepsilon}
\partial_{t_0} \rho + \bm \nabla \cdot (\rho\, \bm u) = 0 \,, \quad \partial_{t_0} (\rho\, \bm u) + \bm \nabla \cdot \bm \Pi^{(0)} = 0 \,,
\ee
\be
\label{eq:ce_secondorderepsilon}
\partial_{t_1} \rho = 0 \,, \quad \partial_{t_1} (\rho\, \bm u) + \left(1 - \f{1}{2 \tau}\right) \bm \nabla \cdot \bm \Pi^{(1)} = 0 \,.
\ee
The tensor $\bm \Pi$ has the components
\be
\begin{aligned}
\Pi_{\alpha \beta} &= \sum_i c_{i \alpha} c_{i \beta} f_i \,, \\
\Pi_{\alpha \beta}^{(0)} &= \sum_i c_{i \alpha} c_{i \beta} f_i^{(0)} \,, \\
\Pi_{\alpha \beta}^{(1)} &= \sum_i c_{i \alpha} c_{i \beta} f_i^{(1)} \,.
\end{aligned}
\ee
$c_{i \alpha}$ denotes the $\alpha$-component of the velocity vector $\bm c_i$. $\Pi_{\alpha \beta}$ will be related to the macroscopic momentum flux tensor in subsection \ref{subsec:shearstress}. The combined form of the two time scales $t_0$ and $t_1$ finally reads
\be
\label{eq:finallbequation}
\begin{aligned}
\partial_t \rho + \bm \nabla \cdot (\rho \bm u) &= 0 \,, \\ \partial_t (\rho\, \bm u) + \bm \nabla \cdot (\rho\, \bm u \bm u) &= - c_s^2 \bm \nabla \rho + 2 \nu \bm \nabla \cdot (\rho \bm S) \,.
\end{aligned}
\ee
$\nu = c_s^2 (\tau - 1 / 2)$ is the kinematic viscosity of the LB fluid. The shear rate tensor $\bm S$ has the components
\be
\label{eq:shearratetensor}
S_{\alpha \beta} = \f 12 \left(\partial_\alpha u_\beta + \partial_\beta u_\alpha\right) \,.
\ee
The incompressible Navier-Stokes equations read
\be
\label{eq:incompnavierstokes}
\bm \nabla \cdot \bm u = 0 \,, \quad \rho\, \partial_t \bm u + \rho\, \bm \nabla \cdot (\bm u \bm u) = - \bm \nabla p + 2 \nu \rho\, \bm \nabla \cdot \bm S \,.
\ee
In the incompressible limit ($\text{Ma} \to 0$ and $\Delta x \to 0$), Eqs.\ (\ref{eq:finallbequation}) and (\ref{eq:incompnavierstokes}) are equivalent, if we assume as equation of state $p = c_s^2 \rho$ for the LB fluid. This means that the LBM asymptotically solves the incompressible Navier-Stokes equations. For a detailed Chapman-Enskog analysis we refer to \cite{hou_simulation_1995}. It is also possible to compute the macroscopic equations of higher orders in $\epsilon$. In \cite{qian_higher-order_2000} the result for $\epsilon^3$ is presented.

As a further comment, we point to non-isothermal lattice models. The standard LBM, as presented in this article, does not satisfy the Galilean invariance principle when considering moments of order higher than two. For the numerically stable modeling of thermal systems, which depend on higher moments, extended lattice structures have to be considered \cite{chikatamarla_entropy_2006, siebert_lattice_2008, nie_galilean_2008}.


\subsection{Shear stress in the lattice Boltzmann method}
\label{subsec:shearstress}

Fluids (liquids and gases alike) are described by the incompressible Navier-Stokes equations, cf.\ Eq.\ (\ref{eq:incompnavierstokes}), in the limit of small Knudsen and Mach numbers. Those equations can also be written in the form $\rho\, \partial_t u_\alpha = - \partial_\beta M_{\alpha \beta}$ with the total momentum flux
\be
\label{eq:momentumflux}
M_{\alpha \beta} = p\, \delta_{\alpha \beta} + \rho\, u_\alpha u_\beta - \sigma_{\alpha \beta} \,,
\ee
where $\delta_{\alpha \beta}$ is the usual Kronecker symbol. The first term on the right-hand-side of Eq.\ (\ref{eq:momentumflux}) is the isotropic pressure and the second one the momentum transfer due to mass transport. The third term describes the momentum diffusion due to the viscosity of the fluid. It is called the deviatoric shear stress. It only vanishes if the velocity gradients are all zero or if the entire fluid is rotating with a spatially constant frequency (as to ensure that no relative motion occurs within the fluid). In a viscous fluid, momentum diffuses from regions with large to regions with small momentum. If the fluid is incompressible, this deviatoric shear stress tensor, from now on only called shear stress tensor, takes a simple form,
\be
\label{eq:shearstresstensor}
\sigma_{\alpha \beta} = \eta \left(\partial_\alpha u_\beta + \partial_\beta u_\alpha\right) \,.
\ee
$\eta = \rho\, \nu$ is the dynamic viscosity. For Newtonian fluids, the shear rate tensor $\bm S$ Eq.\ (\ref{eq:shearratetensor}) and the shear stress tensor $\bm \sigma$ Eq.\ (\ref{eq:shearstresstensor}) are related through $2 \rho\, \nu \bm S = \bm \sigma$. Both tensors are symmetric and traceless. The latter property holds as long as the fluid is incompressible, since $\text{Tr}\, \bm S = \bm \nabla \cdot \bm u$. It is known that some fluids show a shear-thinning or shear-thickening behavior, i.e., the viscosity is a function of the shear rate (e.g., blood or honey). If the fluid is Newtonian, viscosity is not a function of the shear rate, and $\bm S$ and $\bm \sigma$ differ only by a constant factor. Since we restrict ourselves to the latter case, unless otherwise stated, the terms `shear rate' and `shear stress' can be used synonymously.

It follows from the Chapman-Enskog expansion that the momentum flux tensor $\bm M$ from the Navier-Stokes equations can be approximated by the second moments of the LB populations. The equilibrium parts of the populations lead to the pressure and the convective term in the momentum flux tensor,
\be
\label{eq:momentumflux1}
\Pi_{\alpha \beta}^{(0)} = \sum_i c_{i \alpha} c_{i \beta} f_i^{(0)} = c_s^2 \rho\, \delta_{\alpha \beta} + \rho\, u_\alpha u_\beta \,.
\ee
The shear stress is provided by
\be
\label{eq:momentumflux2}
\begin{aligned}
\Pi_{\alpha \beta}^{(1)} &= \sum_i c_{i \alpha} c_{i \beta} f_i^{(1)} \\
&= - \f{\tau}{3} \left[\partial_\alpha (\rho\, u_\beta) + \partial_\beta (\rho\, u_\alpha) + \partial_{t_0} (\rho\, u_\alpha u_\beta)\right] \\
&\approx - \f{\tau}{3 \nu} \sigma_{\alpha \beta} \,.
\end{aligned}
\ee
Obviously, there appear some correction terms in Eq.\ (\ref{eq:momentumflux2}) which need to be discussed. We will come back to this issue in subsection \ref{subsec:convergence}. The total Navier-Stokes momentum flux can be approximated by ($p = c_s^2 \rho$)
\be
\label{eq:totalmomentumflux}
M_{\alpha \beta} = p\, \delta_{\alpha \beta} + \rho\, u_\alpha u_\beta - 2 \rho \nu S_{\alpha \beta} \approx \Pi_{\alpha \beta}^{(0)} + \left(1 - \f{1}{2 \tau}\right) \Pi_{\alpha \beta}^{(1)} \,,
\ee
where $\nu = c_s^2 (\tau - 1 / 2)$ has been used.

At this point a difficulty arises. It is easy to separate the equilibrium populations $f_i^{(0)} = f_i^{\text{eq}}$ from the remaining corrections, the non-equilibrium populations $f_i^{\text{neq}} = f_i - f_i^{\text{eq}}$. But it is not possible to compute the first order populations $f_i^{(1)}$ as stand-alone quantities. This means that $f_i^{\text{neq}}$ has to be used instead of $f_i^{(1)}$. Obviously, this additional approximation only concerns the shear stress tensor. The remaining macroscopic variables (density, velocity) are not influenced. Sometimes the tracelessness of the shear stress tensor is enforced explicitly by writing \cite{mei_force_2002}
\be
\label{eq:shearstress_lb}
\sigma_{\alpha \beta} = - \left(1 - \f{1}{2 \tau}\right) \sum_i \left(c_{i \alpha} c_{i \beta} - \f{\delta_{\alpha \beta}}{D} \bm c_i \cdot \bm c_i\right) f_i^{\text{neq}} \,,
\ee
where $D$ is the number of spatial dimensions. Due to the approximation $f_i^{(1)} \to f_i^{\text{neq}}$ and especially the influence of $f_i^{(2)}$, it is generally expected that the LBM result for the shear stress converges with a first order rate in the case of diffusive scaling ($\Delta t \propto \Delta x^2$) --- in contrast to the velocity, which has a second order accuracy. This convergence behavior is further discussed in subsection \ref{subsec:convergence} and tested in section \ref{sec:simulations}.

Up to this point we have not mentioned the special role of the shear stress in the LB simulations: Spatial derivatives on a lattice are usually computed by using finite difference (FD) schemes. The advantage of the LBM is that the shear stress --- although related to the gradient of the fluid velocity --- can directly be assessed locally, i.e., on each individual lattice node, and independently of the velocity. No information of neighbors is required, and the implementation is straightforward. This is the strength of the LBM. Nevertheless, in order to have a comparison, we have examined in section \ref{sec:simulations} both the shear stress obtained from the local approach given in Eq.\ (\ref{eq:shearstress_lb}) and a non-local finite difference method. All finite differences in this paper are of second-order accuracy and central, e.g.,
\be
\f{\partial u_x}{\partial y} (x, y, z) = \f{u_x (x, y + 1, z) - u_x (x, y - 1, z)}{2} \,.
\ee


\subsection{Convergence of the shear stress}
\label{subsec:convergence}

Knowing the Navier-Stokes equations and the macroscopic limit of the LB equation, the basic idea is to estimate the error of the solution obtained by the LBM, as compared to the Navier-Stokes solution. As also mentioned in the introduction, in the diffusive scaling, the relative error of the velocity is proportional to $\text{Ma}^2$ (see e.g., \cite{holdych_truncation_2004} and references therein). The same is true for the density fluctuations, i.e., the compressibility of the LB fluid also scales with $\text{Ma}^2$. The convergence behavior of the shear stress, on the other hand, is not obvious and has to be deduced from the LB equation.

Since the equilibrium populations can be computed exactly (in terms of the Maxwell truncated approximation), the equilibrium part $\Pi_{\alpha \beta}^{(0)}$ Eq.\ (\ref{eq:momentumflux1}) contains only compressibility and truncation errors of the same order as the velocity and the density. Higher order terms of $f_i$ do not play a role, since they do not appear in the equation. In the diffusive scaling, the relative velocity and density errors are $e_u \propto \text{Ma}^2$ and $e_\rho \propto \text{Ma}^2$ and, for this reason, the relative error of $\Pi_{\alpha \beta}^{(0)}$ also scales with $\text{Ma}^2$.

Things are different for the shear stress $\Pi_{\alpha \beta}^{(1)}$ in Eq.\ (\ref{eq:momentumflux2}). We write the LBM shear stress as the Navier-Stokes shear stress plus an error term,
\be
\Pi_{\alpha \beta}^{(1)} = - \f{\tau}{3} \rho \left(\partial_\alpha u_\beta + \partial_\beta u_\alpha\right) + E_{\Pi^1} \,.
\ee
It can be shown \cite{hou_simulation_1995} that this error term reads
\be
\label{eq:ce_pierror}
E_{\Pi^1} = - \tau \left(\rho\, u_\alpha u_\gamma \partial_\gamma u_\beta + \rho\, u_\beta u_\gamma \partial_\gamma u_\alpha + u_\alpha u_\beta \partial_\gamma (\rho\, u_\gamma)\right) \,.
\ee
The point here is that not only the errors of $\rho$ and $u_\alpha$ account for the total deviation of the shear stress, but also the additional terms in $E_{\Pi^1}$. For the relative error it follows that $e_{\Pi^1} = E_{\Pi^1} / \Pi_{\alpha \beta}^{(1)} \propto \text{Ma}^2$, since the three relations $u_\alpha \propto \text{Ma}$, $\partial_\alpha \propto \text{Ma}$, and $\partial_\alpha \rho \propto \text{Ma}^3$ hold. The first statement is obvious. The second one can be understood by recalling the diffusive scaling $\Delta x \propto \text{Ma}$. When the grid is refined and $\Delta x$ decreased, all spatial derivatives \textit{in lattice units} are also decreased by the same rate, i.e., $\partial_\alpha \propto 1 / N \propto \Delta x$. The last statement follows from the second one and the fact that the LB fluid compressibility scales with $\text{Ma}^2$, i.e., the fluctuations of the density $\rho$ with respect to the mean density $\bar \rho$ behave like $(\rho - \bar \rho) / \bar \rho \propto \text{Ma}^2$.

This indicates that the shear stress would converge with a second order rate, \textit{if} the populations $f_i^{(1)}$ were known and used for its computation. As we have stated before, it is not possible to evaluate $f_i^{(1)}$. In principle, the populations $f_i^{(1)}$ can be computed from the equilibrium via Eq.\ (\ref{eq:ce_firstorder}). This is not practical, because information from next neighbors and different time steps would be necessary in order to evaluate this equation. The LBM obviously would loose the advantage of locality, and the shear stress could directly be computed using a finite difference scheme for the equation $\sigma_{\alpha \beta} = \nu \rho (\partial_\alpha u_\beta + \partial_\beta u_\alpha)$ in the first place, without bothering computing $f_i^{(1)}$ at first and $\Pi_{\alpha \beta}^{(1)}$ afterwards. Hence, the first order terms $f_i^{(1)}$ are substituted by the non-equilibrium populations in the simulations and $\Pi_{\alpha \beta}^{\text{neq}}$ is computed instead of $\Pi_{\alpha \beta}^{(1)}$.

As a consequence, the influence of the higher order corrections $f_i^{(>1)}$ plays a role. Fortunately, the higher order populations become less important with increasing order of $\epsilon$. Hence, we focus only on the influence of $f_i^{(2)}$ and check whether its contribution may destroy the second order convergence of the shear stress. One can write
\be
\label{eq:noneqstress}
\Pi_{\alpha \beta}^{\text{neq}} = - \f{\tau}{3} \rho \left(\partial_\alpha u_\beta + \partial_\beta u_\alpha\right) + E_{\Pi^1} + E_\epsilon \,.
\ee
$E_\epsilon$ is the error due to the higher order corrections, especially $f_i^{(2)}$. Thus, we are interested in the convergence behavior of $\Pi_{\alpha \beta}^{(2)} = \sum_i c_{i \alpha} c_{i \beta} f_i^{(2)}$. From the Chapman-Enskog analysis we know the population $f_i^{(2)}$, cf.\ Eq.\ (\ref{eq:ce_secondorder}), and thus
\be
\label{eq:secondorder}
- \f{1}{\tau} \Pi_{\alpha \beta}^{(2)} = \partial_{t_1} \Pi_{\alpha \beta}^{(0)} + \left(1 - \f{1}{2 \tau}\right) \left(\partial_{t_0} \Pi_{\alpha \beta}^{(1)} + \partial_\gamma R_{\alpha \beta \gamma}^{(1)}\right) \,,
\ee
where $R_{\alpha \beta \gamma} = \sum_i c_{i \alpha} c_{i \beta} c_{i \gamma} f_i$ is the third moment. From Eq.\ (\ref{eq:ce_firstorder}) it follows
\be
R_{\alpha \beta \gamma}^{(1)} = - \tau \left(\partial_{t_0} R_{\alpha \beta \gamma}^{(0)} + \partial_\zeta P_{\alpha \beta \gamma \zeta}^{(0)}\right)
\ee
with the fourth moment $P_{\alpha \beta \gamma \zeta}^{(0)} = \sum_i c_{i \alpha} c_{i \beta} c_{i \gamma} c_{i \zeta} f_i^{(0)}$. Since the standard equilibrium Eq.\ (\ref{eq:equilibrium}) is truncated after the second order in the velocity, all velocity moments larger than the second order are not physically well defined. Indeed, in the framework of the truncated equilibrium, all equilibrium moments larger than two scale with $\rho\, u^2 \propto \text{Ma}^2$.

At this point we have to admit that it is a tedious task to exactly compute $\bm \Pi^{(2)}$. Actually, this is not necessary, because the focus is put on the convergence behavior and thus the scaling with the Mach number. The following considerations are straightforward if one recalls that $u_\alpha \propto \text{Ma}$, $\partial_\alpha \propto \text{Ma}$, and $\partial_\alpha \rho \propto \text{Ma}^3$: From the above observations we conclude that all terms in Eq.\ (\ref{eq:secondorder}) are of order $\text{Ma}^4$. Additionally, Eqs.\ (\ref{eq:ce_firstorderepsilon}), (\ref{eq:ce_secondorderepsilon}), (\ref{eq:momentumflux1}), and (\ref{eq:momentumflux2}) have been used. This means that both error terms $E_{\Pi^1}$ and $E_\epsilon$ are of fourth order in the Mach number, whereas $\bm \Pi^{(1)}$ itself is second order. As a consequence, the relative error of the deviatoric shear stress, as computed from Eq.\ (\ref{eq:noneqstress}), is quadratic---similar to the errors of the velocity and the density. This is a quite encouraging result. However, as will be shown below, the second order convergence is very sensitive to additional error sources. In particular, our simulations clearly show that the convergence behavior strongly depends on the type of boundary conditions.

In subsection \ref{subsec:results_convergence}, our numerical results are presented and compared to the above qualitative findings.


\subsection{Geometry and boundary conditions}
\label{subsec:geometry}

For the simulations in this article we have chosen a simple 3D geometry, the laminar 3D Poiseuille flow in a channel with rectangular cross section. The fluid enters the numerical grid at $x = 1$ and exits at $x = N_x$. The channel has height $H$ (along $z$-axis), width $W$ (along $y$-axis), and length $L = (N_x - 1) \Delta x$. The Reynolds and Mach numbers are
\be
\label{eq:reynoldsmach}
\text{Re} = \f{H \hat u}{\nu} \,, \quad \text{Ma} = \f{\hat u}{c_s} \,.
\ee
$\hat u$ is the center velocity of the flow.

The velocity field for an infinitely long channel with rectangular cross section in 3D is known analytically. The Navier-Stokes equations reduce to the Poisson-like equation $\eta\, \Delta u_x (y, z) = \partial p / \partial x = \text{const}$. It can be solved by expanding the solution in basis functions. One possible solution \cite{haberman_applied_2004} is
\label{eq:velocity1}
\begin{multline}
\shoveleft{\makebox[\width]{$\displaystyle u_x (y, z) = \f{\hat u_x}{\Sigma} \bigg[z (H - z)$}} \\
 - \f{8 H^2}{\pi^3} \sum_n^{\text{odd}} \f{1}{n^3} \f{\cosh(n \pi (y - W / 2) / H)}{\cosh(n \pi W / 2 H)} \sin \left(\f{n \pi z}{H}\right)\bigg]
\end{multline}
with the origin of the coordinate system at one corner of the channel. The pressure gradient has been eliminated in favor of the center velocity $\hat u_x$, $\partial p / \partial x = - 2 \eta \hat u_x / \Sigma$, where $\Sigma$ is a normalization factor depending on the geometry. For the case of the rectangular channel described above,
\be
\label{eq:sigma}
\Sigma = \f{H^2}{4} - \f{8 H^2}{\pi^3} \sum_n^{\text{odd}} \f{1}{n^3} \f{\sin (n \pi / 2)}{\cosh(n \pi W / 2 H)} \propto H^2\,.
\ee
Another equivalent solution \cite{haberman_applied_2004} is
\be
\label{eq:velocity2}
u_x (y, z) = \f{\hat u_x}{\Sigma} \f{32}{\pi^4} \sum_{n,m}^{\text{odd}} \f{1}{n m} \f{\sin \f{n \pi y}{W} \sin \f{m \pi z}{H}}{(n / W)^2 + (m / H)^2}
\ee
with the same value of $\Sigma$.

Both solutions have advantages and disadvantages. The first solution has only one expansion coefficient and converges faster. However, while convergence is fast in the region near $y = W / 2$, the no-slip condition at $y = 0$ and $y = W$ is poorly reproduced. The second solution, on the other hand, shows very good convergence near the entire border, since the basis functions vanish there. However, two expansion coefficients are necessary. Moreover, the convergence is slower when approaching the center of the channel. Therefore, when computing the analytic solution, we divide the system into an inner part, where Eq.\ (\ref{eq:velocity1}) is used, and the wall region, where the exact solution is computed via Eq.\ (\ref{eq:velocity2}). The shear rate tensor $\bm S$ can directly be calculated from both solutions.

The accuracy of the simulation is quantified using the common L2-norm for the relative error
\be
\label{eq:l2error}
e_u = \sqrt{\f{\sum |u_{\text{an}} - u_{\text{sim}}|^2}{\sum |u_{\text{an}}|^2}} \,,
\ee
where the sum takes into account all lattice nodes in the numerical grid. The error of the two relevant shear components $S_{xy}$ and $S_{xz}$ are defined similarly.

The no-slip condition at the side walls is realized by a fullway bounce-back rule. The walls are located halfway between the fluid and the wall nodes, and the accuracy is of second order \cite{he_analytic_1997}. For the inlet and outlet we have used periodic boundary conditions in one part of the simulations and velocity boundary conditions in the other part. In the case of periodic boundary conditions, the flow is driven by a constant and homogeneous body force \cite{guo_discrete_2002}. This force can be implemented by adding $w_i\, \bm c_i \cdot \bm F / c_s^2$ to the right-hand-side of Eq.\ (\ref{eq:lbequation}), where $\bm F$ is the body force pointing in $x$-direction.

We have examined both the velocity boundary conditions proposed by Skordos \cite{skordos_initial_1993} and by Latt \cite{latt_straight_2008}. The velocity profiles on the inlet and outlet are fully developed. Implementing velocity boundary conditions is not as straightforward as using the bounce-back rule. The underlying idea is the following: If the fluid velocity $\bm u$ is given on a straight wall, the populations $f_i$ have to be computed from $\bm u$. This is in general a problem, since the number of equations available is smaller than the number of unknown populations. For this reason, some closure relations are assumed, which complete the set of equations. There are various possibilities to do this. A good review of the velocity boundary conditions can be found in \cite{latt_straight_2008}. Using an explicit method, it boils down to the equation
\be
f_i = f_i^{\text{eq}} (\rho, \bm u) + f_i^{(1)} \,, \quad f_i^{(1)} = - \f{w_i \tau \rho}{c_s^2} Q_{i \alpha \beta} S_{\alpha \beta}
\ee
with $Q_{i \alpha \beta} = c_{i \alpha} c_{i \beta} - c_s^2 \delta_{\alpha \beta}$. Following this general approach, the boundary conditions also show a second order convergence. This is important in order not to spoil the quality of the overall simulation. It depends on the method how $f_i^{(1)}$ is estimated. Skordos uses a finite difference method for the velocity gradient, while Latt approximates the unknown values of $f_i^{(1)}$ with the help of the known ones. The density $\rho$ at the inlet and outlet can be recovered by
\be
\label{eq:densityonboundaries}
\rho = \f{1}{1 + u_\perp} (2 \rho_+ + \rho_0) \,,
\ee
where $u_\perp$ is the projection of the boundary velocity $\bm u$ on the boundary normal, pointing outside the numerical grid, i.e., $u_\perp$ is negative at the inlet and positive at the outlet. This way, the correct pressure gradient is recovered. $\rho_+$ is the sum of all populations leaving the numerical grid (those populations are known), and $\rho_0$ is the sum of all populations streaming parallel to the boundary plane (those are also known).


\subsection{Influence of the relaxation parameter}
\label{subsec:relparameter}

The accuracy of a LB simulation depends on the LBM itself, the initial conditions, the boundary conditions and also on the simulation parameters. There are four relevant quantities which have to be set up for every simulation: The Reynolds number $\text{Re}$ which is fixed by the physical system, the Mach number $\text{Ma}$ of the simulation, the dimensionless relaxation parameter $\tau$ and the lattice constant $\Delta x$ or, equivalently, the number $N$ of fluid lattice nodes along one axis (we use $N_z$). Using Reynolds and Mach numbers from Eq.\ (\ref{eq:reynoldsmach}) yields an important relation between those four dimensionless parameters,
\be
\label{eq:lbquantities}
\f{\text{Ma}}{\text{Re}} = \f{1}{\sqrt{3}} \f{\tau - \f 12}{N_z} \,.
\ee
Assuming that the Reynolds number is fixed, there are two degrees of freedom in the choice of the parameters. Although the Mach number in reality is uniquely defined by the speed of sound, it can be varied in the LBM. As long as the Mach number is small, $\text{Ma} < {\cal O} (0.1)$, it does not influence the physics of the system. This way, in a simulation the value of $\text{Ma}$ can be adjusted in order to increase the accuracy or shorten the computing time.

The second order convergence of the LB equation can only be realized, if the Mach number and the lattice constant are decreased simultaneously. The diffusive scaling fulfills this requirement \cite{reider_accuracy_1995, holdych_truncation_2004}, and it is ensured that the incompressible Navier-Stokes equations are recovered in the limit of $\Delta x \to 0$. The disadvantage of this method is that by reducing the Mach number the required number of time steps is increased (recall that $\Delta t \propto \Delta x^2$). As a consequence, halving the lattice constant and the Mach number simultaneously reduces the error by a factor of $2^2$ but increases the simulation time by a factor of $2^4$ (2D) or $2^5$ (3D).

Equation (\ref{eq:lbquantities}) plays an important role by the selection of simulation parameters. In order to see this, we note that the choice of the Reynolds number and the Mach number is closely linked to the physical problem of interest. In the diffusive scaling, on the other hand, the Mach number uniquely sets the lattice resolution $\Delta x$ and the time step $\Delta t$ (recall that $\Delta x \propto \text{Ma}$ and $\Delta t\propto \text{Ma}^2$). Thus, for fixed values of $\text{Re}$ and $\text{Ma}$, the only freedom is in a variation of the relaxation parameter $\tau$ or, equivalently, the number of grid points $N_z$.

In many cases of interest, however, the Reynolds number and the Mach number can be varied in a reasonable range without changing the basic physics of the problem. When studying problems, where sound propagation does not play a role, it is often possible to increase the Mach number as long as density fluctuations remain negligibly small. This brings the advantage of using larger grid resolutions and hence significantly reducing the computation time. A similar philosophy also applies to the choice of the Reynolds number: When studying laminar flow problems, $\text{Re}$ can be increased as long as velocity fluctuations remain negligibly small. A more complete description of these and similar ideas can be found in \cite{cates_physical_2005}.

Therefore, the question arises which initial choice of the four parameters appearing in Eq.\ \ref{eq:lbquantities} is optimal, meaning which choice leads to the smallest possible error at constant simulation time. There is a very detailed analysis of the LB error in \cite{holdych_truncation_2004}. The authors point out that the accuracy of the LBM strongly depends on the choice of the relaxation parameter $\tau$. This can also be seen in \cite{qian_higher-order_2000}. The important point is that the error has a minimum for $\tau$ in a region between $0.8$ and $1.0$, depending on the Reynolds number. Setting $\tau$ too close to $1 / 2$ or larger than $1$ results in avoidable inaccuracies of the simulations.

In order to verify the results presented in \cite{holdych_truncation_2004} and to see whether the velocity error and the shear stress error behave similarly we have set up a series of simulation according to the following paradigm:
\begin{enumerate}
\item Set $\text{Re}$ to the desired value.
\item Choose a reasonable value for $\text{Ma}$.
\item There is only one free parameter left. Set up a series of simulations for different grid sizes $N$. This automatically fixes $\tau$ according to Eq.\ (\ref{eq:lbquantities}). It is wise to start for $\tau$ around $0.9$.
\item Calculate the relative errors $e_u$ and $e_S$ of the simulations. The curves $e_u (\tau)$ and $e_S (\tau)$ both feature a minimum, which can be found by varying $N$ and thus $\tau$.
\item The minimum value $\tau_{\text{opt}}$ is the best choice for the simulations. It should be kept fixed, if the resolution is to be changed. This directly leads to the diffusive scaling and a second order convergence.
\end{enumerate}

There are some remarks: In the simulations, the minimum value of $\tau$ is not necessarily the minimum value of the LBM, because the accuracy of the boundary conditions also depends on $\tau$ and --- in general --- in another way than the LBM. The bounce-back rule for example introduces a slip velocity at the walls, which depends on the relaxation parameter \cite{he_analytic_1997}. Additionally, the quality of the velocity boundary conditions depends on the Mach number. For this reason, the location of the optimum value $\tau_{\text{opt}}$ is also a function of $\text{Ma}$ for larger Mach numbers.

We present our results for $\tau_{\text{opt}}$ in subsection \ref{subsec:results_initialparameters}.


\section{Mass increase}
\label{sec:massincrease}

Some wall boundary conditions do not obey local mass conservation. This shortcoming can be removed by applying modified boundary conditions \cite{chopard_mass_2003, hollis_enhanced_2006}. But even if the mass is conserved at the walls locally, it is possible that the total mass of the numerical grid changes with time \cite{chopard_mass_2003, sukop_lattice_2005}. This effect is caused by inlet and outlet velocity boundary conditions. Due to the equation of state of the lattice fluid, $p = c_s^2 \rho$, a pressure gradient is equivalent to a density gradient, leading to a larger inlet than outlet flux, and mass is accumulating in the system. There are possibilities to oppose this effect, e.g., the use of velocity inlet / pressure outlet boundary conditions or incompressible LB models, both for steady and unsteady flows \cite{he_lattice_1997, guo_lattice_2000}. Mass increase is fully absent also in the case of a body force driven flow with periodic boundary conditions. However, we are interested in the consequences of the mass increase, if it is not avoided in the first place.

The reader should be reminded that $L$, $W$ and $H$ are the length, width and height of the channel, whereas $N_x$, $N_y$ and $N_z$ are the number of fluid lattice nodes along the length, width and height of the channel, respectively. In lattice units, obviously $L = N_x - 1$, $W = N_y$ and $H = N_z$ hold, as long as inlet / outlet are defined by velocity boundary conditions and the side walls by the bounce-back rule. The side walls themselves are not contained in $N_y$ and $N_z$, so the total lattice --- including fluid and walls --- has dimensions $N_x \times (N_y + 2) \times (N_z + 2)$.

For a 3D rectangular Poiseuille flow, the velocity profiles at the inlet and the outlet have to be equal, $u = u_{\text{in}} = u_{\text{out}}$, resulting in a net total mass increase per time step
\be
\label{eq:massincrease}
\dot M := \f{\Delta M}{\Delta t} = - \Delta \rho\, A\, \bar u > 0 \,,
\ee
where $\Delta \rho$ is the density difference between the outlet and the inlet (and thus negative), $A$ is the cross section area and $\bar u$ the mean velocity on the cross sections. Note that $\bar u$ and $\hat u$ are proportional to the Mach number and that $\hat u / \bar u$ is constant for a given channel geometry. In a 2D Poiseuille flow one finds $\hat u / \bar u = 3 / 2$. In three dimensions, on the other hand, this ratio depends on the aspect ratio $W / H$ of the cross section.

In an ideal LB simulation of Poiseuille flow along the $x$-axis, the density is only a function of $x$. Furthermore only the velocity component $u_x$ does not vanish. It is constant along the $x$-axis and depends on $y$ and $z$. In this case, the Stokes equation for the LB fluid reads (assuming that steady state is reached, i.e., $\partial_t \bm u = 0$)
\be
\f{\partial \ln \rho}{\partial x} = - \f{\nu \nabla^2 u}{c_s^2} = - \f{6 \nu \hat u}{\Sigma} = \text{const} \,,
\ee
whose integrated form reads
\be
\label{eq:densitygradient}
\rho = \rho_0\, \exp \left(- \f{6 \nu \hat u}{\Sigma} x\right) \,.
\ee
$\Sigma$ is the normalization factor from Eq.\ (\ref{eq:sigma}). The pressure difference between inlet ($x = 1$) and outlet ($x = N_x$) is
\be
\Delta \rho = \rho_0 \left[\exp \left(- \f{6 \nu \hat u}{\Sigma} N_x\right) - \exp \left(- \f{6 \nu \hat u}{\Sigma}\right)\right] \,.
\ee
The inlet and outlet densities are given only implicitly by Eq.\ (\ref{eq:densityonboundaries}). It is, therefore, not possible to set the velocity profiles \textit{and} the densities on the inlet / outlet simultaneously. For small values of $6 \nu \hat u N_x / \Sigma$, the density gradient can be approximated by the expression from subsection \ref{subsec:geometry}: $- 6 \rho_0 \nu \hat u L / \Sigma$, i.e.,
\be
\f{\Delta p}{L} = - \f{2 \rho_0 \nu \hat u}{\Sigma} \,.
\ee
This is valid whenever the Mach number is small and the density near $\rho_0$ everywhere.

The total mass in the numerical grid can be estimated by
\be
\begin{aligned}
M &\approx \int dV\, \rho = A \int_{1/2}^{N_x+1/2} dx\, \rho \\
&= - A \rho_0 \f{\Sigma}{6 \nu \hat u} \exp \left(- \f{3 \nu \hat u}{\Sigma}\right) \left[\exp \left(- \f{6 \nu \hat u}{\Sigma} N_x\right) - 1\right] \,.
\end{aligned}
\ee
Putting everything together results in the expression
\be
\f{\dot M}{M} = \f{6 \nu \bar u \hat u}{\Sigma} \f{1 - \exp \left(\f{6 \nu \hat u}{\Sigma} (N_x - 1)\right)}{1 - \exp \left(\f{6 \nu \hat u}{\Sigma} N_x\right)} \exp \left(\f{3 \nu \hat u}{\Sigma}\right) \approx \f{6 \nu \bar u \hat u}{\Sigma} \,.
\ee
The approximation in the second step is justified if $N_x$ and $\Sigma$ are sufficiently large, which is always the case for an appropriate lattice resolution.

Since the geometry and the Reynolds number shall be fixed here, the Mach number can only be adjusted by changing the kinematic viscosity $\nu$ according to $\nu \propto \text{Ma}$ so that finally
\be
\label{eq:massfunction}
M (N_t) = M_0\, \exp \left(C_g \text{Ma}^3 N_t / \text{Re}\right) \,.
\ee
The mass after $N_t$ time steps depends on the Mach and Reynolds numbers and also on a geometry factor
\be
\label{eq:geometryfactor}
C_g = \f{2}{\sqrt 3} \f{H}{\Sigma} \f{\bar u}{\hat u} \propto \f 1H \,,
\ee
which is uniquely fixed by the shape of the channel. The Mach number enters the equation with a power of $3$. It is an observation that usually $C_g \text{Ma}^3 / \text{Re} \ll 1$, but for a large number $N_t$ of time steps the mass increase may enter its exponential regime. When examining the validity of Eq.\ (\ref{eq:massfunction}), one must pay attention to the following aspects:
\begin{enumerate}
\item The equation is not valid at the very beginning of the simulations, but only when a `steady state' has been reached in which the pressure gradient is fully developed. Here, steady state means that the velocity does not change in time, although the density is not constant. For this reason, the first iteration steps are neglected when comparing simulation results with the analytic expression Eq.\ (\ref{eq:massfunction}).
\item Modifying the Mach number over a wide range, but keeping $\text{Re}$ and $N$ fixed requires changing $\tau$ over a wide range. One has to be careful that $\tau$ stays in a region, where the boundary conditions and the LBM work reliably.
\end{enumerate}
We will show in section \ref{sec:simulations} that once the above considerations are taken into account, Eq.\ (\ref{eq:massfunction}) produces excellent results.

Another question is how a change of the simulation parameters influences the mass increase. If $\text{Re}$ is changed, this has to be compensated by varying the Mach number, the grid resolution or the kinematic viscosity or any combination of those. Suppose the following independent transformations: $\text{Re} \to a\, \text{Re}$, $\text{Ma} \to b\, \text{Ma}$ and $N_z \to c\, N_z$. Since $C_g \propto 1 / N_z$, $N_t \propto N_z^2 / \nu$, $1 / \nu = \text{Re} / (N_z \h u)$ and $1 / \h u \propto 1 / \text{Ma}$, one finds that $C_g \text{Ma}^3 N_t \to b^2\, C_g \text{Ma}^3 N_t$. For fixed aspect ratio of the channel dimensions, this means that the mass increase is directly related to the Mach number and is independent of the other parameters. The compressibility (controlled by the Mach number) and the mass increase are tightly connected.

The mass increase actually raises a problem: The fluid velocity within the LBM is defined in such a way that a global variation of the mass does not change its value, as long as the pressure gradient is unchanged and the distribution of the populations is affected only by a velocity-independent prefactor, cf. Eq.\ (\ref{eq:lbmmomenta}). The definition of the shear stress in Eq.\ (\ref{eq:shearstress_lb}) on the other hand is sensitive to the global mass, since $\bm \sigma \propto \rho$. If the non-physical mass increase is significant, the shear stress tensor must be corrected for this effect.

There are in principal two options for this correction. First, one may take measures to compensate the mass increase in the first place. Second, one may allow the mass increase, but make sure that all observables remain unaffected. We assume that the shear \textit{rate} tensor Eq.\ (\ref{eq:shearratetensor}) is not affected by the mass increase, since it does not depend on the density. Hence it is reasonable to calculate this quantity instead of the shear stress tensor. Due to $2 \nu \rho\, \bm S = \bm \sigma$, one finds, starting from Eq.\ (\ref{eq:shearstress_lb}),
\be
\label{eq:shearrate}
S_{\alpha \beta} = - \f{3}{2 \tau \rho} \sum_i \left(c_{i \alpha} c_{i \beta} - \f{\delta_{\alpha \beta}}{3} \bm c_i \cdot \bm c_i\right) f_i^{\text{neq}} \,.
\ee
In order to check whether this approach is reasonable, we have computed the relative error $e_S$ of the shear rate as a function of time in a `steady state' simulation with significant mass increase. The results are discussed in subsection \ref{subsec:results_massincrease}.


\begin{figure}
\includegraphics[width=8cm,clip=true]{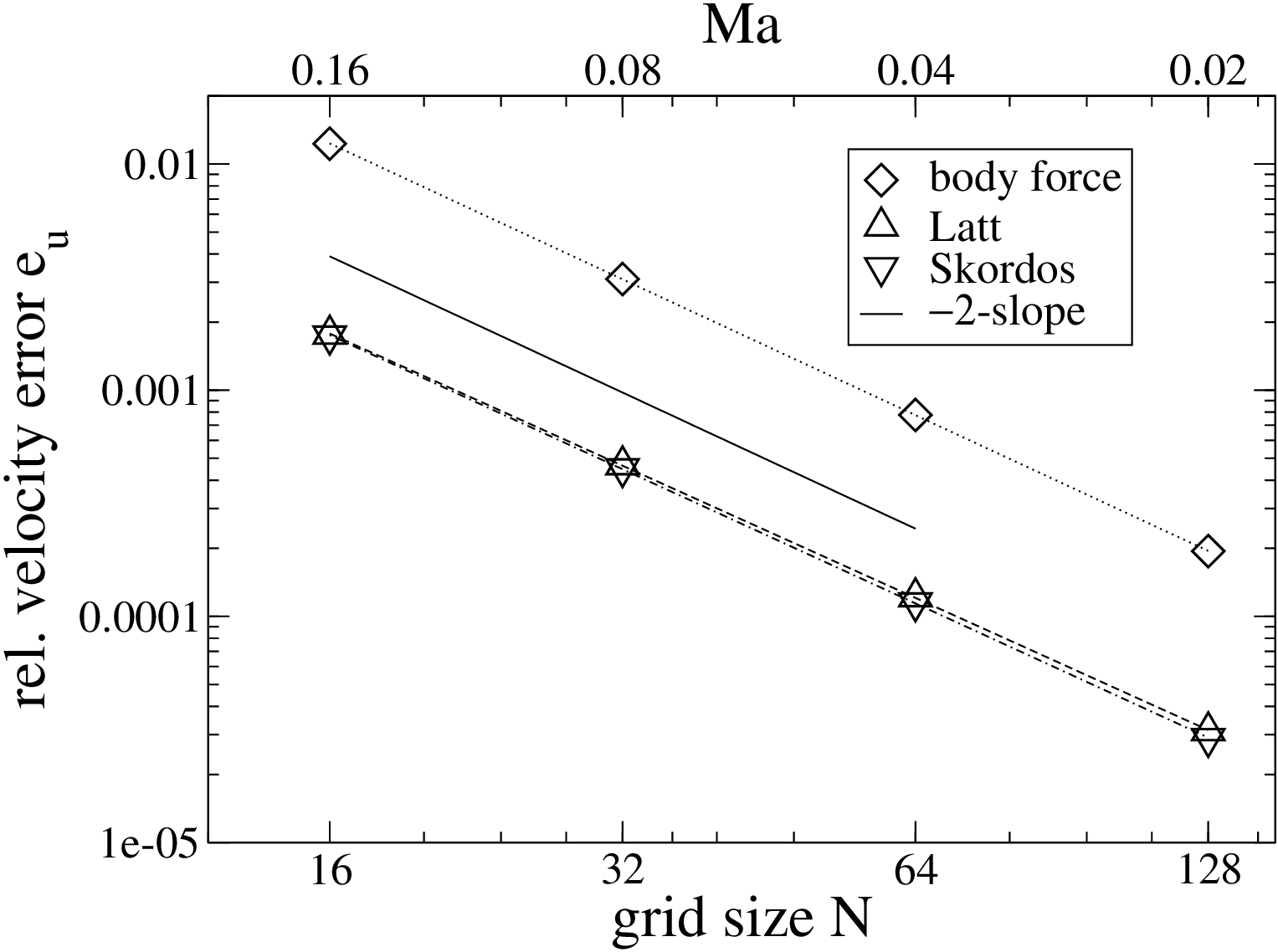} \\[2ex]
\includegraphics[width=8cm,clip=true]{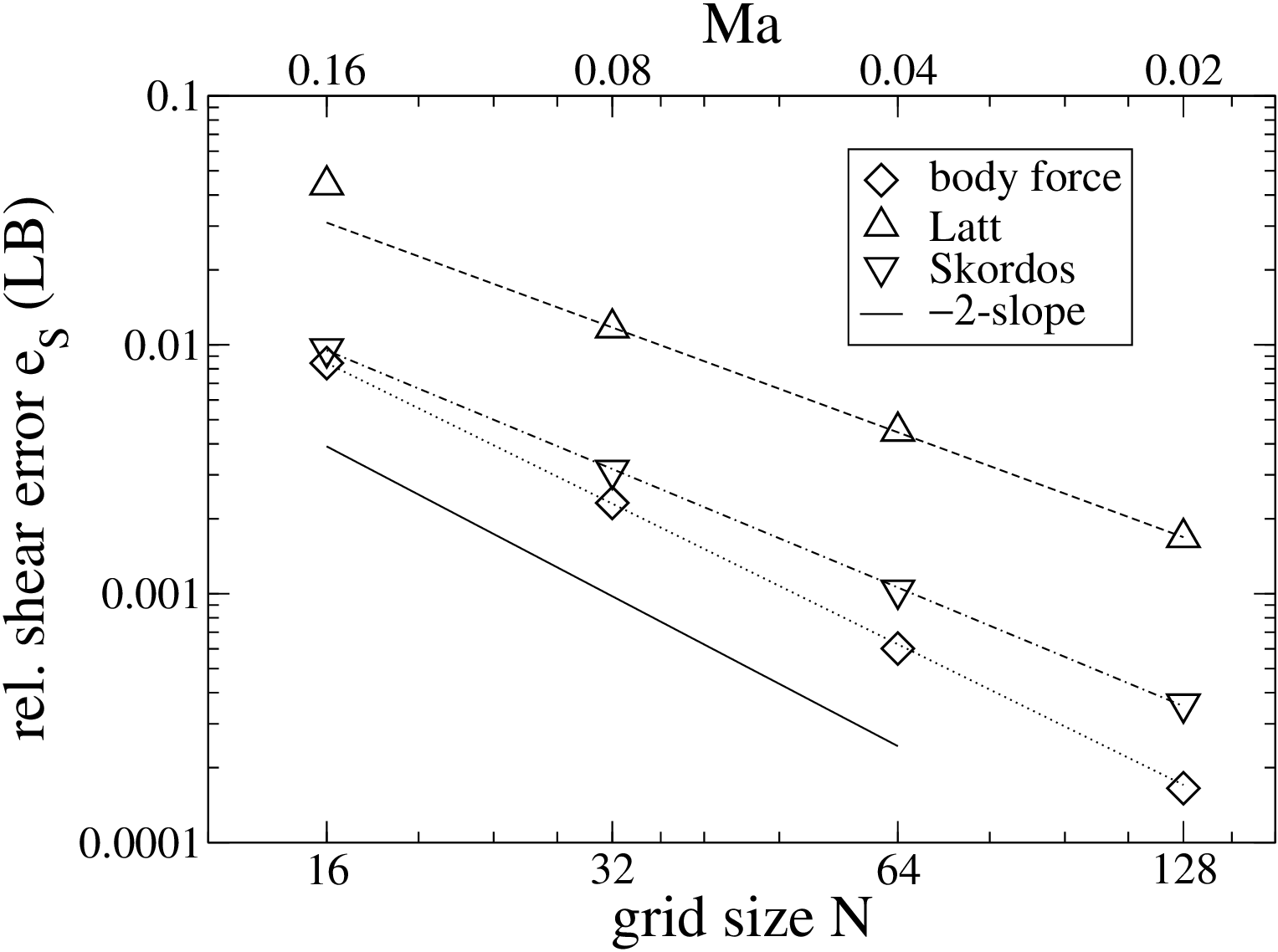} \\[2ex]
\includegraphics[width=8cm,clip=true]{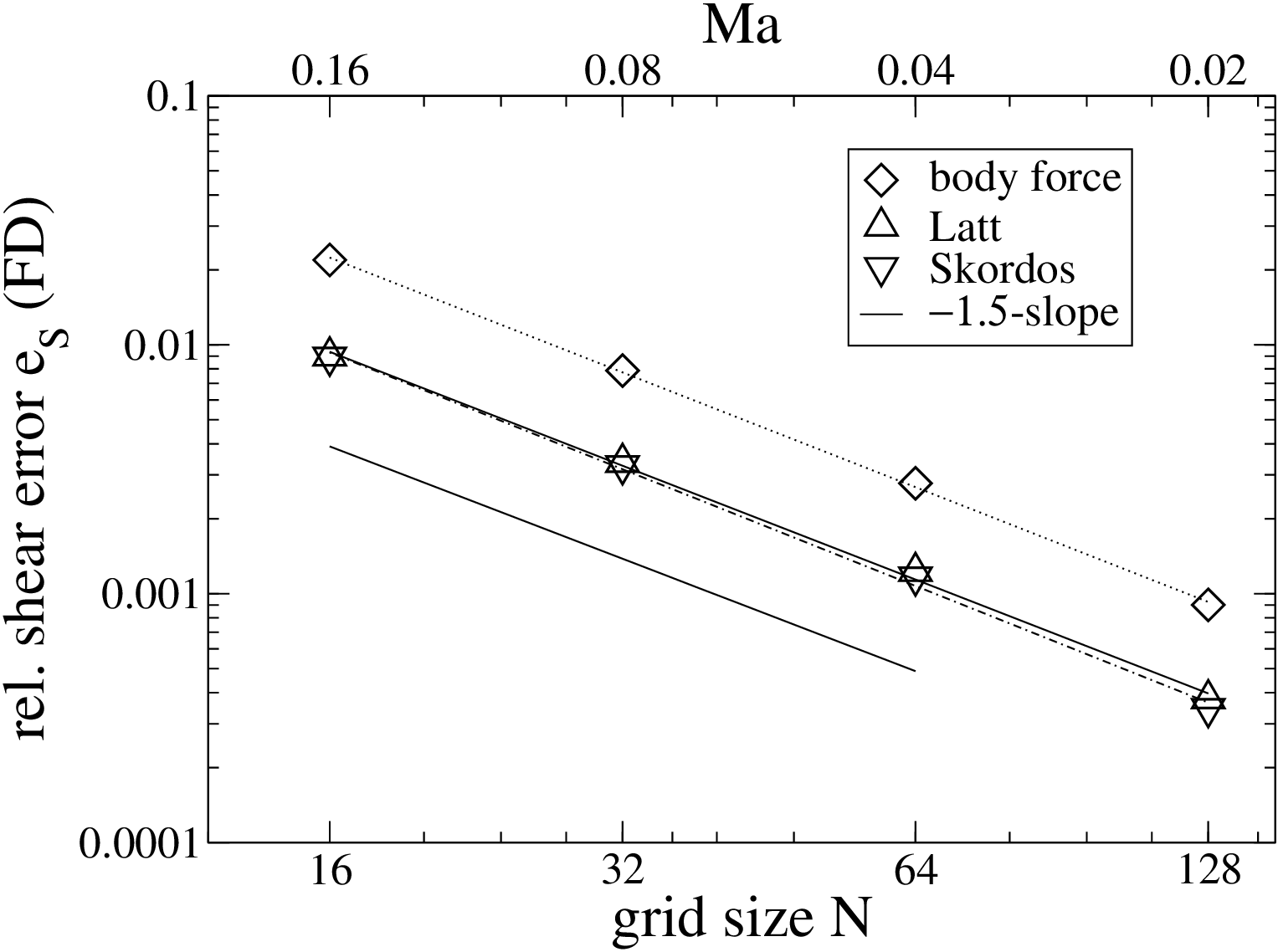}
\caption{Relative L2 errors of the velocity $u_x$ and the shear stress component $S_{xy}$ (both local and FD) as functions of the dimensionless system size $N$: The velocity slope is close to $-2$ for all boundary conditions, whereas the shear stress slope is between $-1.4$ and $-1.9$ [local via Eq.\ (\ref{eq:shearrate})] and around $-1.5$ (FD). The slope values are collected in table \ref{tab:slopes}.}
\label{fig:convergence}
\end{figure}

\begin{figure*}
\includegraphics[width=8cm,clip=true]{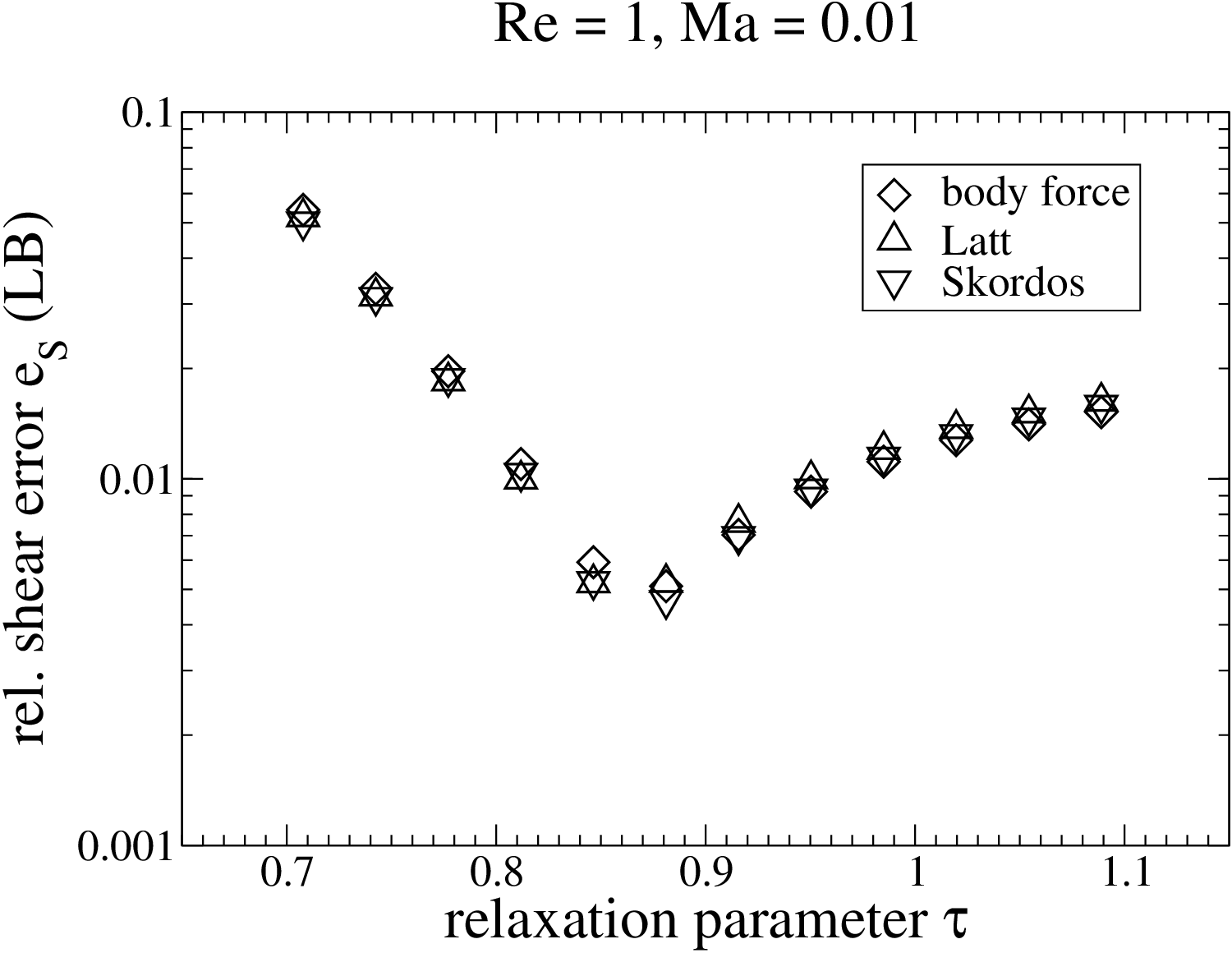} \hspace{1ex}
\includegraphics[width=8cm,clip=true]{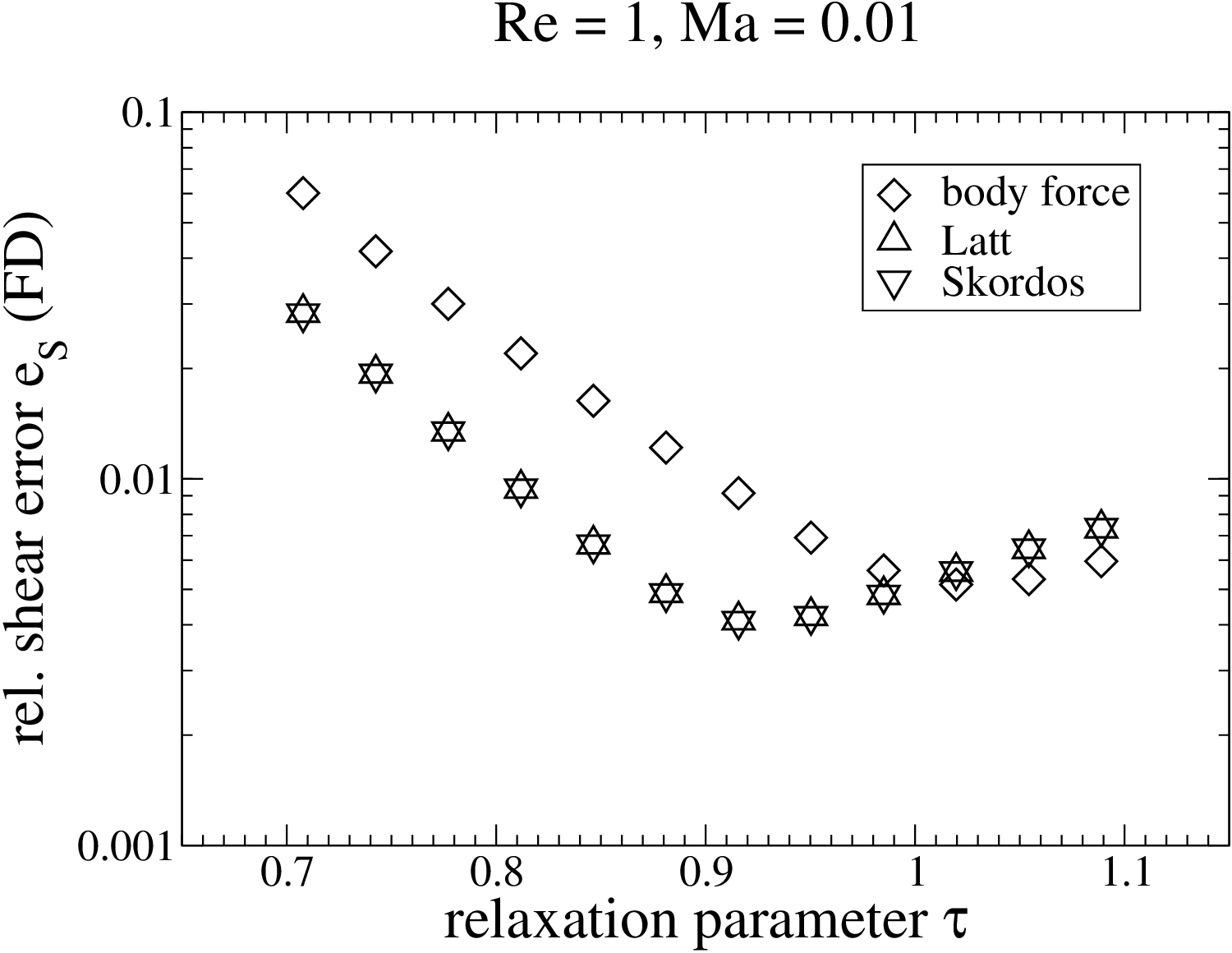} \\ \vspace{2ex}
\includegraphics[width=8cm,clip=true]{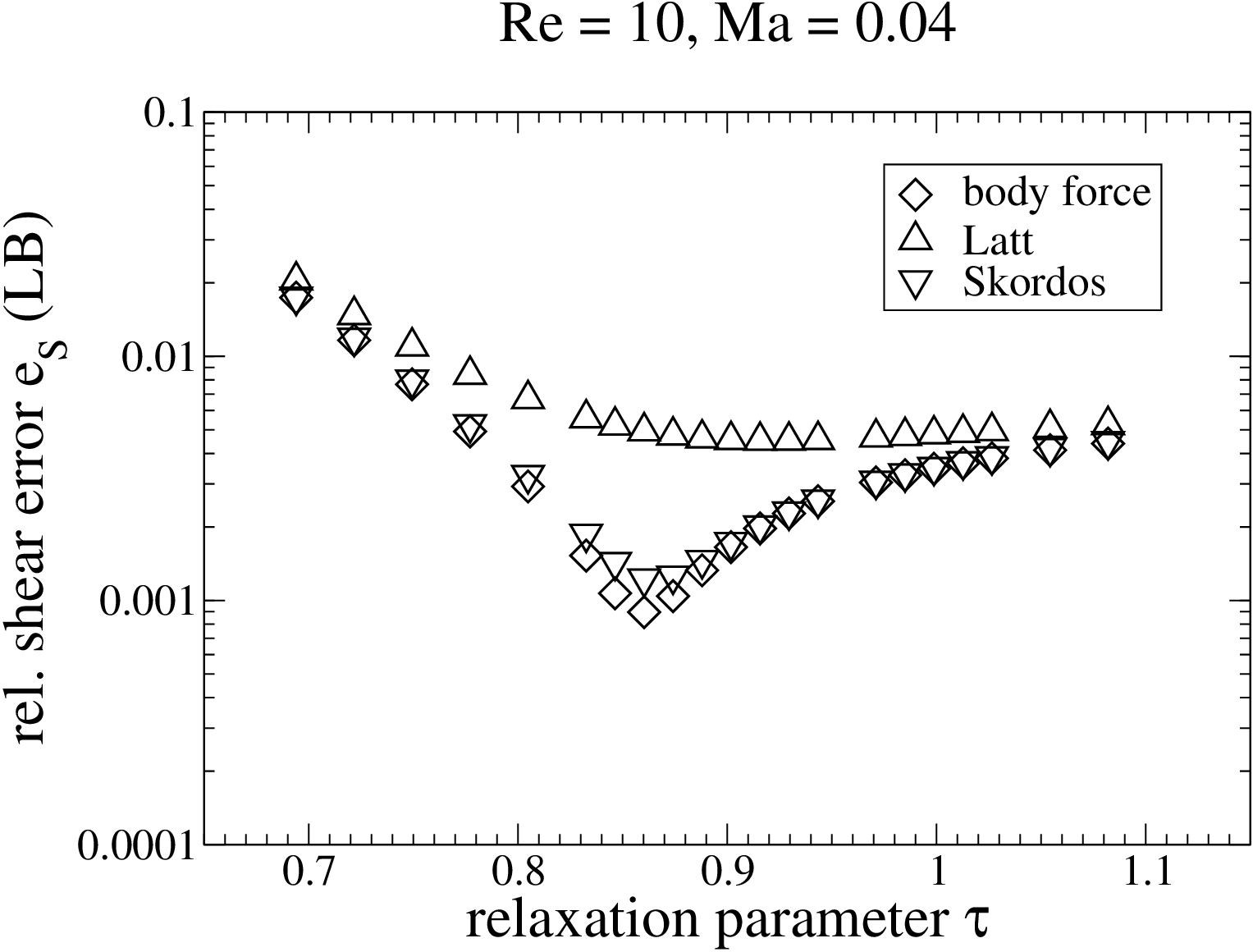} \hspace{1ex}
\includegraphics[width=8cm,clip=true]{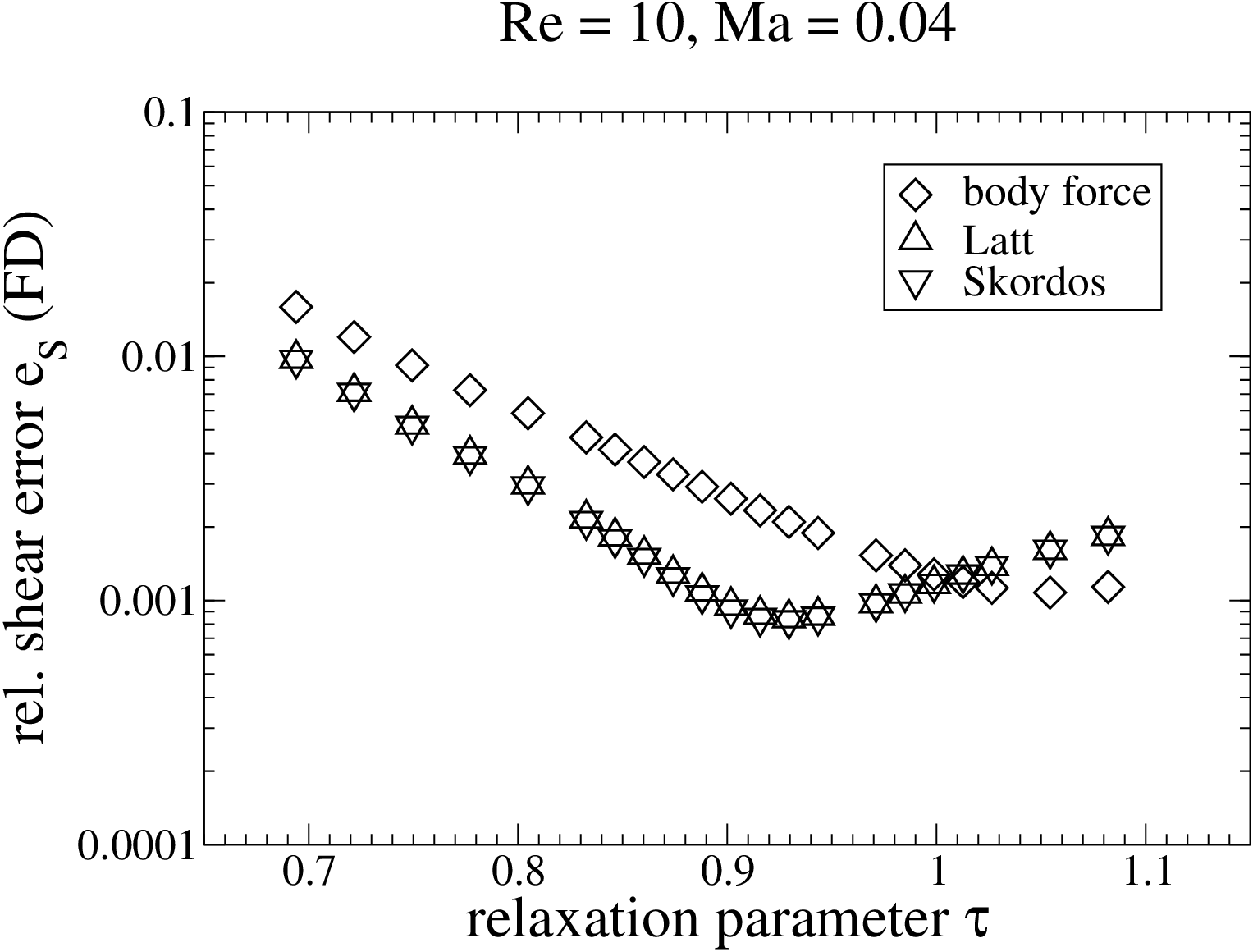} \\ \vspace{2ex}
\includegraphics[width=8cm,clip=true]{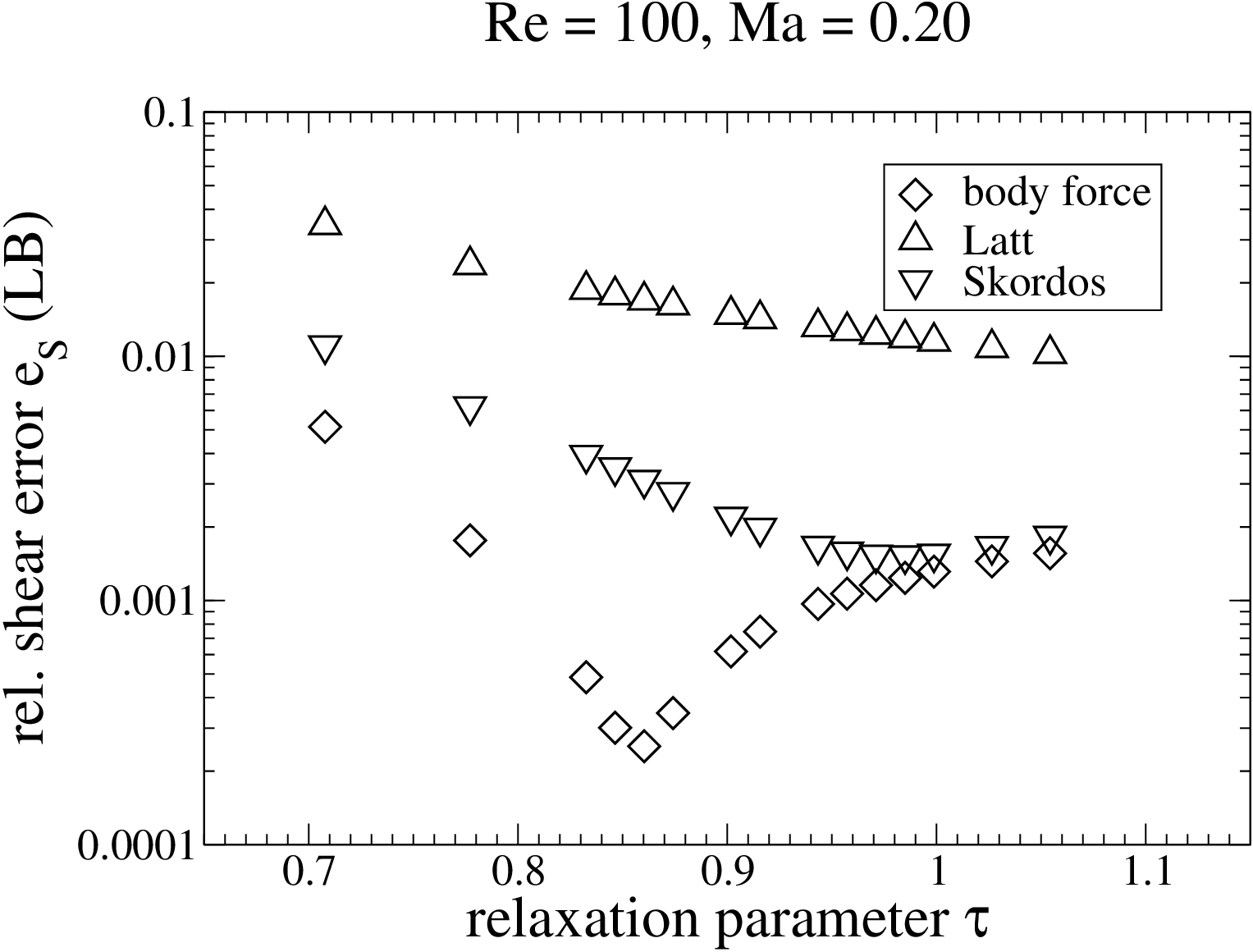} \hspace{1ex}
\includegraphics[width=8cm,clip=true]{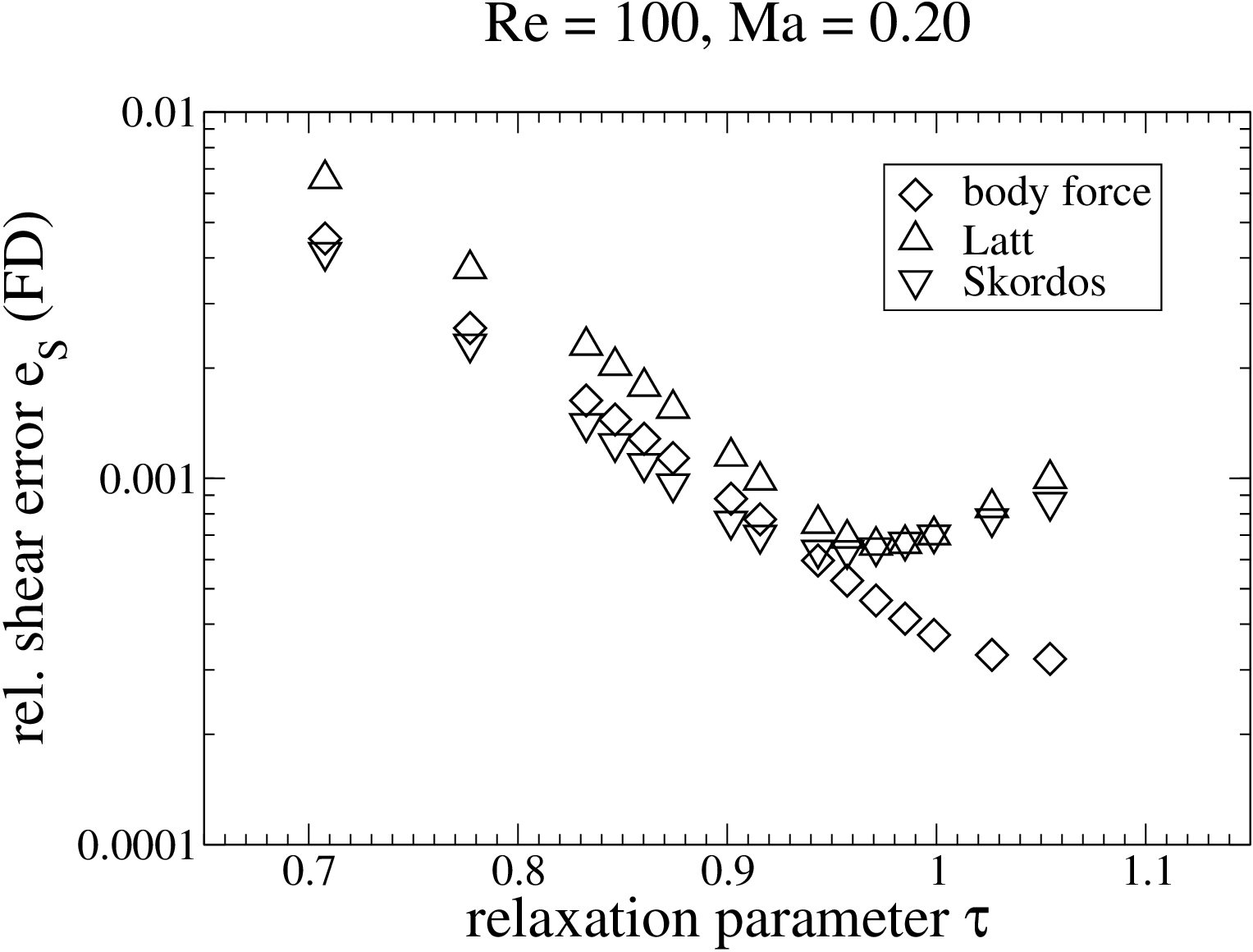}
\caption{Behavior of the relative shear error $e_S (\tau)$ [left column: from Eq.\ (\ref{eq:shearrate}), right column: from FD] for Reynolds numbers $1$ (top), $10$ (middle), $100$ (bottom).}
\label{fig:minimumtau_shear}
\end{figure*}

\section{Simulations and results}
\label{sec:simulations}

\begin{figure}
\includegraphics[width=8cm,clip=true]{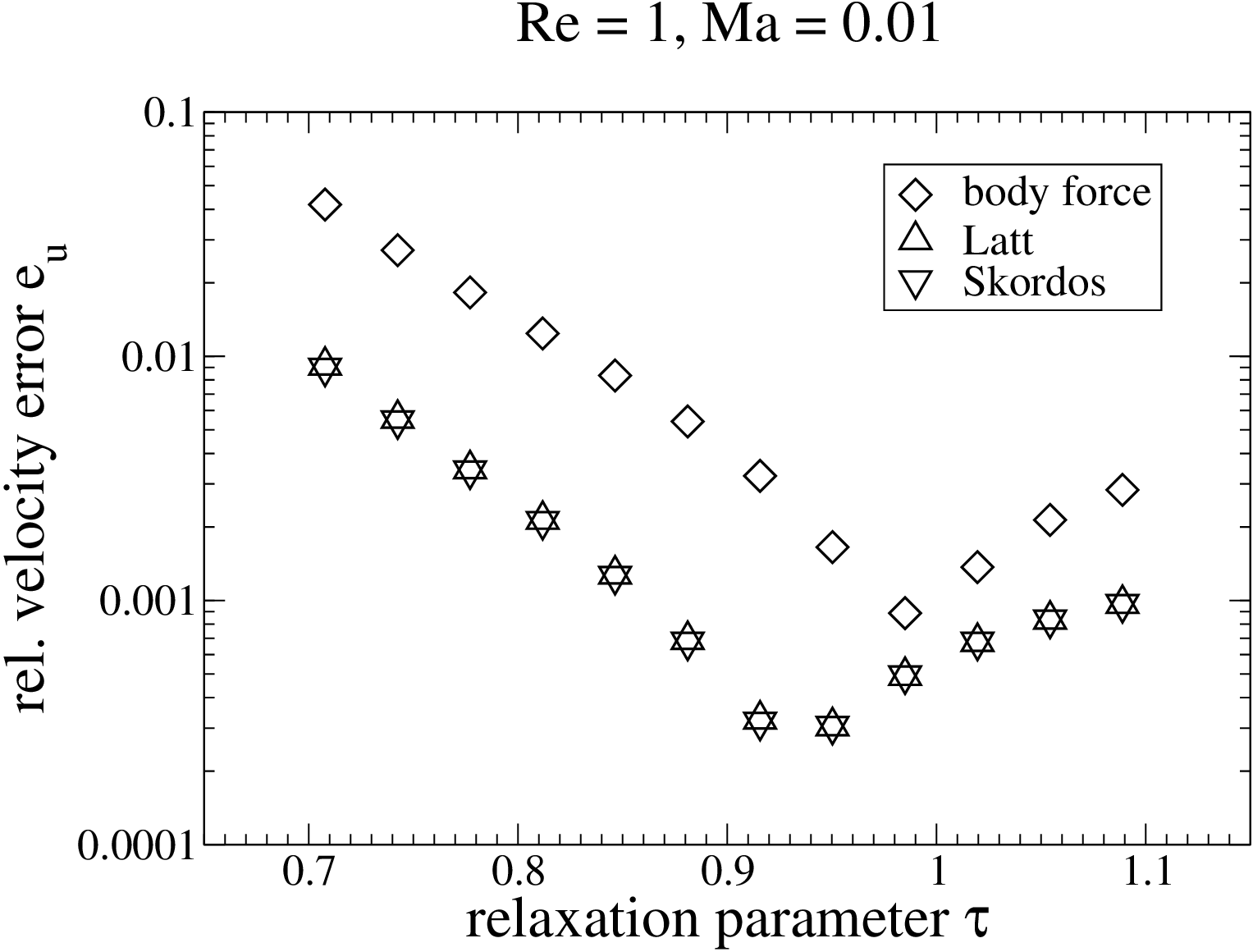} \\ \vspace{2ex}
\includegraphics[width=8cm,clip=true]{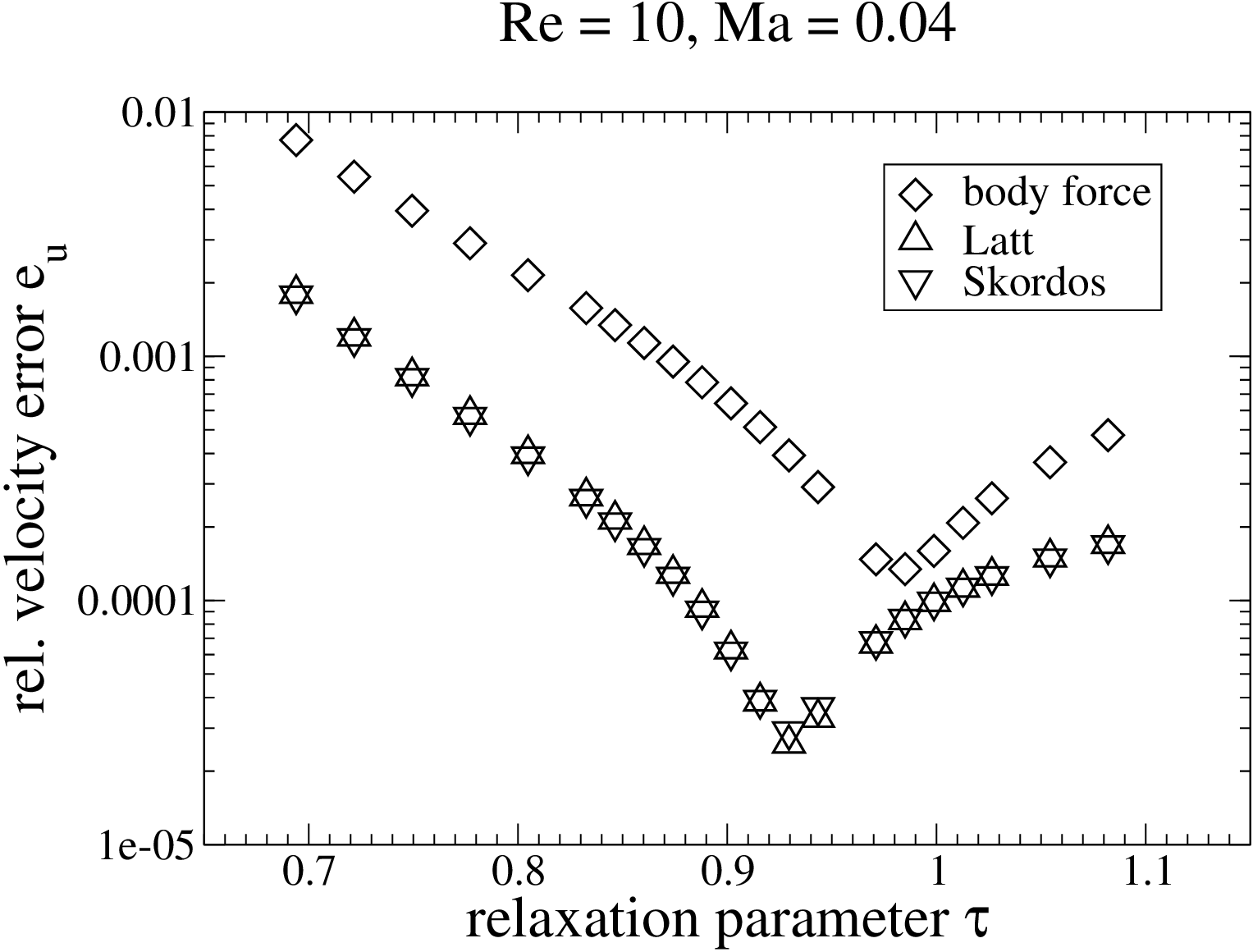} \\ \vspace{2ex}
\includegraphics[width=8cm,clip=true]{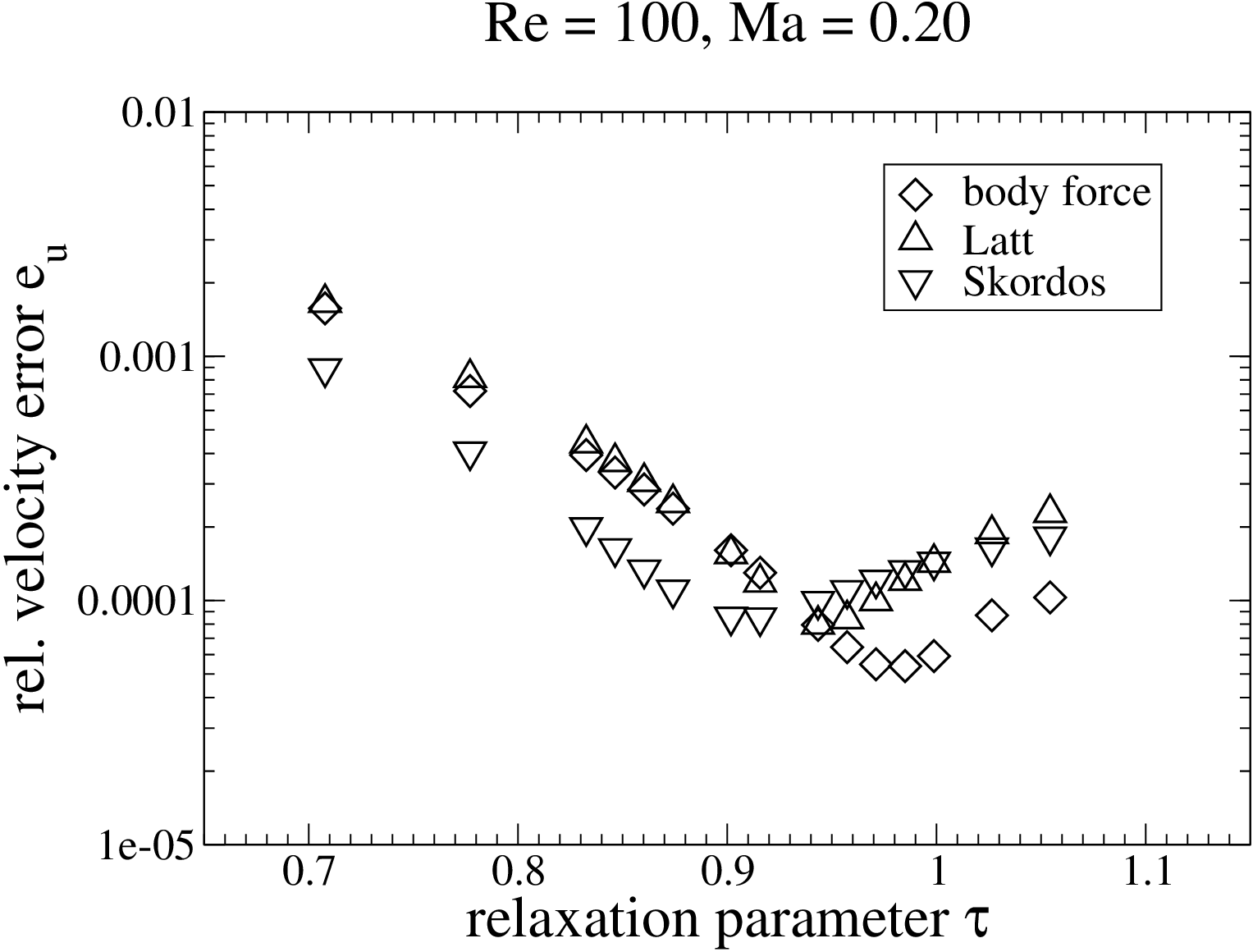}
\caption{Behavior of the relative velocity error $e_u (\tau)$ for Reynolds numbers $1$ (top), $10$ (middle), $100$ (bottom).}
\label{fig:minimumtau_vel}
\end{figure}

We investigate the velocity and shear stress error behavior in a laminar 3D Poiseuille flow using both periodic boundary conditions with body force and velocity boundary conditions. We are interested in the convergence of the numerical results towards the analytical solutions. Moreover, we determine the optimum choice of the relaxation parameter, depending on whether the velocity or the shear stress is the relevant observable. Those results are presented in subsections \ref{subsec:results_convergence} and \ref{subsec:results_initialparameters}

The computing time of a LB simulation can be reduced if the Mach number is increased. The reason is that less time steps are necessary to reach the final state of the flow. This is very important whenever large Reynolds numbers are given, because simulations with high Reynolds number require a large numerical grid. However, the increase of the Mach number results in a larger numerical error due to compressibility effects and, in combination with velocity boundary conditions, it leads to the violation of mass conservation in the system. This mass increase in a steady Poiseuille flow is examined qualitatively and quantitatively in subsection \ref{subsec:results_massincrease}.

The presented benchmark analyses are not intended to and cannot cover the above mentioned aspects in every detail. Rather, our aim is a fundamental understanding of the discussed observations. Nevertheless, it would be an interesting task for future work to extend the above studies to other situations such as complex geometries as well as non-steady flows, issues not considered in this work.

\subsection{Convergence}
\label{subsec:results_convergence}

\begin{table}
\caption{\label{tab:slopes} Slope values of the velocity $u_x$ and shear rate $S_{xy}$ errors for the three boundary conditions (BC). The corresponding figures are presented in Fig.\ \ref{fig:convergence}.}
\begin{ruledtabular}
\begin{tabular}{lccc}
BC & $u_x$ & $S_{xy}$ (local) & $S_{xy}$ (FD) \\ \hline
periodic & $-1.99$ & $-1.88$ & $-1.49$ \\
Latt & $-1.94$ & $-1.40$ & $-1.52$ \\
Skordos & $-1.97$ & $-1.62$ & $-1.56$ \\
\end{tabular}
\end{ruledtabular}
\end{table}

In subsection \ref{subsec:convergence}, we have argued that the convergence of the shear stress should in principle have a $\Delta x^{-2}$- or $\text{Ma}^{-2}$-behavior. This prediction has been tested with a series of simulations at fixed $\text{Re} = 12.5$ and $\tau \approx 0.855$. The size of the numerical boxes has been $16 \times 18 \times 18$ up to $128 \times 130 \times 130$. The errors of the velocity and the shear component $S_{xy}$ are shown in figure \ref{fig:convergence} as a function of the lattice resolution.

There are three relevant observations. First, the relative error $e_u$ of the velocity is significantly smaller than that of the shear rate. This is no surprise, since there are many error terms in Eq.\ (\ref{eq:noneqstress}).

The second and most interesting observation is the slope of the convergence. For the velocity, it is found to be very close to $-2$ for all boundary conditions, see table \ref{tab:slopes}. It shows that the LBM and the boundary conditions are second order accurate with respect to the velocity. This has been shown many times in the literature. For the shear rate Eq.\ (\ref{eq:shearrate}) on the other hand, we have found slopes strongly depending on the boundary conditions. The slope for the periodic boundary condition is $-1.88$ and nearly reaches the predicted value of $-2$. The velocity boundary conditions have flatter slopes, cf.\ table \ref{tab:slopes}, but are clearly better than first order convergence.

Our interpretation is that indeed the shear stress is a second-order accurate observable, but the convergence is detrimentally influenced if velocity boundary conditions are used. Since the implementations of those conditions usually are only self-consistent on the $\epsilon$- and $\epsilon^2$-level, additional errors appear affecting the shear stress.

Another interesting point is that the error of the shear stress obtained from the finite difference scheme is not significantly smaller than that of the local scheme. The slopes for the finite difference shear rate are around $-1.5$ for all investigated boundary conditions.

Let us briefly compare the simulations: The velocity boundary conditions yield a much better accuracy for the velocity field than the periodic boundary condition (nearly one order of magnitude). For the shear stress, things are different: The periodic boundary condition with a body force gives rise to the highest accuracy and the best convergence, but Skordos' method is not significantly worse. Latt's velocity boundary condition is less accurate, but a $1\%$ accuracy can be found even at small resolutions. Additional information on the accuracy of the shear stress is provided in Sec.\ \ref{subsec:results_initialparameters}.

Some remarks have to be made here. Steady Poiseuille flow has the simplest possible geometry. The error of the shear stress is given by the terms $E_{\Pi^1}$ and $E_\epsilon$ in Eq.\ (\ref{eq:noneqstress}). It is expected that those quantities can be much larger in arbitrary flow geometries and non-steady flows, although the second order convergence should be preserved. The small relative errors $e_S = 0.0001 \ldots 0.01$ which have been recovered in these benchmark tests may be a consequence of the high symmetry of the Poiseuille flow. For this reason, more benchmark tests for other, more complex geometries, for which the analytical solution is known, should be performed. Practically, the second order rate is not assumed to be recovered as long as velocity boundary conditions are used.

It is an open question whether there exists a global slip velocity in our simulations, which could be subtracted from the velocity profiles. If the slip velocity is constant over the cross section of the flow, i.e., if the slip velocity is the same at edges and corners, it would be important to identify its value and subtract it from the velocity field. Such a correction is not included in the present work.

\subsection{Initial parameters}
\label{subsec:results_initialparameters}

\begin{table*}
\caption{\label{tab:optimumvalues}Overview of the locations $\tau_{\text{opt}}$ and the relative errors $e(\tau_{\text{opt}})$ of the minima for different boundary conditions (BC) and Reynolds / Mach numbers.}
\begin{ruledtabular}
\begin{tabular}{rclcccccc}
\multirow{2}{*}{$\text{Re}$} & \multirow{2}{*}{$\text{Ma}$} & \multirow{2}{*}{BC} & \multicolumn{2}{c}{velocity} & \multicolumn{2}{c}{shear (LB)} & \multicolumn{2}{c}{shear (FD)} \\
 & & & $\tau_{\text{opt}}$ & $e (\tau_{\text{opt}})$ & $\tau_{\text{opt}}$ & $e (\tau_{\text{opt}})$ & $\tau_{\text{opt}}$ & $e (\tau_{\text{opt}})$ \\ \hline
\multirow{3}{*}{$1$} & \multirow{3}{*}{$0.01$} & periodic & $0.99$ & $9 \cdot 10^{-4}$ & $0.87$ & $5 \cdot 10^{-3}$ & $1.02$ & $5 \cdot 10^{-3}$ \\
 & & Latt & $0.94$ & $3 \cdot 10^{-4}$ & $0.86$ & $5 \cdot 10^{-3}$ & $0.92$ & $4 \cdot 10^{-3}$ \\
 & & Skordos & $0.94$ & $3 \cdot 10^{-4}$ & $0.87$ & $5 \cdot 10^{-3}$ & $0.92$ & $4 \cdot 10^{-3}$ \\ \hline
\multirow{3}{*}{$10$} & \multirow{3}{*}{$0.04$} & periodic & $0.98$ & $1 \cdot 10^{-4}$ & $0.86$ & $9 \cdot 10^{-4}$ & $1.05$ & $1 \cdot 10^{-3}$ \\
 & & Latt & $0.93$ & $3 \cdot 10^{-5}$ & $0.92$ & $5 \cdot 10^{-3}$ & $0.93$ & $8 \cdot 10^{-4}$ \\
 & & Skordos & $0.93$ & $3 \cdot 10^{-5}$ & $0.86$ & $1 \cdot 10^{-3}$ & $0.93$ & $8 \cdot 10^{-4}$ \\ \hline
\multirow{3}{*}{$100$} & \multirow{3}{*}{$0.20$} & periodic & $0.98$ & $5 \cdot 10^{-5}$ & $0.86$ & $3 \cdot 10^{-4}$ & NA & NA \\
 & & Latt & $0.94$ & $8 \cdot 10^{-5}$ & NA & NA & $0.97$ & $7 \cdot 10^{-4}$ \\
 & & Skordos & $0.91$ & $9 \cdot 10^{-5}$ & $0.99$ & $2 \cdot 10^{-3}$ & $0.95$ & $6 \cdot 10^{-4}$ \\
\end{tabular}
\end{ruledtabular}
\end{table*}

For the Reynolds numbers $1$, $10$ and $100$, the optimum values of the relaxation parameter $\tau$ have been investigated according to the procedure described in subsection \ref{subsec:relparameter}. The error curves $e (\tau)$ are depicted in figures \ref{fig:minimumtau_shear} and \ref{fig:minimumtau_vel}, and the optimum values along with the minimum errors are given in table \ref{tab:optimumvalues}.

From a theoretical point of view, $\tau_{\text{opt}}$ depends on the value of the Reynolds number \cite{holdych_truncation_2004}. The analysis in \cite{holdych_truncation_2004} however does only account for the error due to the LBM. The behavior of the boundary conditions is not included. Those effects are clearly visible in our results.

For the velocity in figure \ref{fig:minimumtau_vel} we have found that velocity boundary conditions are in principle more accurate than a body force driven flow. At larger Reynolds numbers, however, when the Mach number is increased in order to shorten the simulation time, the velocity boundary conditions become more inaccurate. But even at the comparably large value $\text{Ma} = 0.20$, the accuracy of all three methods is similar. It strikes the reader that the relaxation parameter should be somewhere between $0.9$ and $1.0$ in order to minimize the error $e_u$. The exact value depends on the boundary condition and it will most likely also depend on the geometry and the Mach number (which we have not tested). In table \ref{tab:optimumvalues}, the locations of the minima are given.

The behavior of the relative shear error $e_S$ is presented in figure \ref{fig:minimumtau_shear}. Both methods (shear locally from Eq.\ (\ref{eq:shearrate}) and a finite difference scheme) can directly be compared. The local shear error is smallest for the body force driven flow, especially at larger Mach numbers. This observation matches the previous discussion in subsection \ref{subsec:results_convergence}, where it is argued that the velocity boundary conditions give rise to additional errors. Both velocity boundary conditions behave differently. Skordos' method is more accurate than Latt's scheme. While the former boundary condition produces errors similar to the body force driven flow at $\text{Ma} = 0.04$, the latter method begins to become inaccurate. Finally, at $\text{Ma} = 0.20$, the error obtained from Latt's method is nearly two orders of magnitude larger than that of the body force driven flow, and also Skordos' methods becomes less accurate. The position of the minimum error is at $\tau < 0.9$ for the reasonable simulations and becomes larger when the boundary conditions are inaccurate. The reason is that large $\tau$ is equivalent to high grid resolution (or small $\Delta x$) thus reducing the error related to a higher Mach number. From the plots in figure \ref{fig:minimumtau_shear} we learn that the accuracy of the shear suffers from velocity boundary conditions at relatively large Mach numbers.

This shortcoming can be avoided if a finite difference scheme for the shear is used. Here, the relative error is tightly connected to the error of the velocity, since the shear is interpolated from the velocity information. It may be surprising that the minimum error of the shear from a finite difference scheme is not significantly smaller than that of the shear obtained from Eq. (\ref{eq:shearrate}). So there is no a priori advantage of using a finite difference scheme, but --- at larger Mach numbers --- the finite difference scheme is clearly superior, at least for the velocity boundary conditions.

In conclusion we briefly summarize the most important observations:
\begin{enumerate}
\item The minimum error is always located somewhere around the interval $[0.8, 1.1]$ for the relaxation parameter $\tau$, depending on which quantity is computed and which method is used for this. It is therefore very reasonable to use a relaxation parameter out of this interval for simulations. $\tau \approx 0.9$ is a compromise for most simulations.
\item For the velocity field, velocity boundary conditions are more accurate up to $\text{Ma} = 0.20$.
\item One has to act with caution, if the error of the shear shall be minimized. Velocity boundary conditions become increasingly inaccurate for large Mach numbers, if Eq.\ (\ref{eq:shearrate}) is used. A finite difference scheme overcomes this disadvantage, but the implementation is more difficult.
\item Keeping the Mach number small in general is recommended, but not always practical in large Reynolds number simulations. If a simulation requires a large Reynolds number, one should decrease $\tau$ and increase $\text{Ma}$ reasonably, keeping in mind which observable ($\bm u$ or $\bm \sigma$) is of most interest.
\end{enumerate}

\begin{figure}
\includegraphics[width=8cm,clip=true]{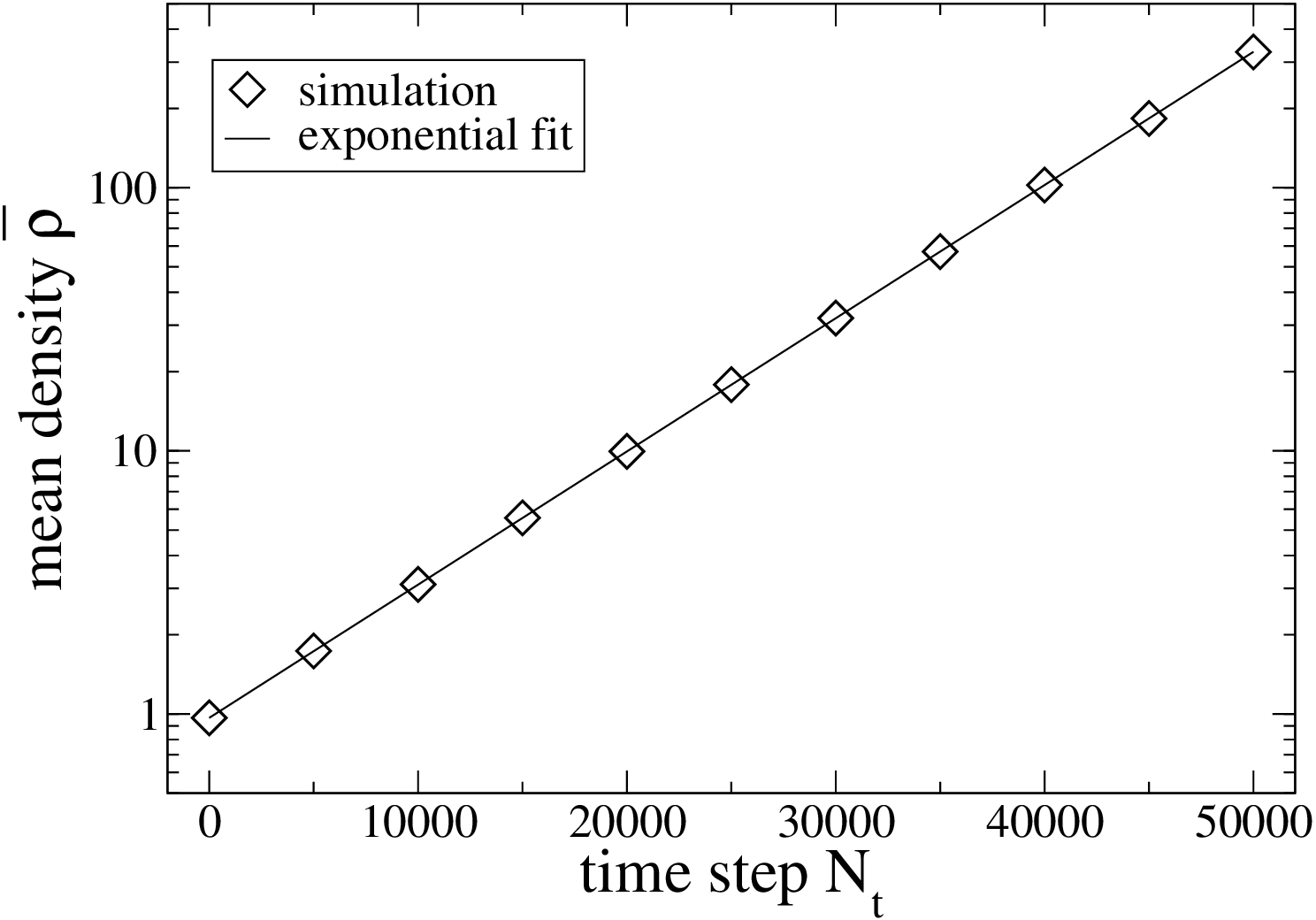} \\ \vspace{2ex}
\includegraphics[width=8cm,clip=true]{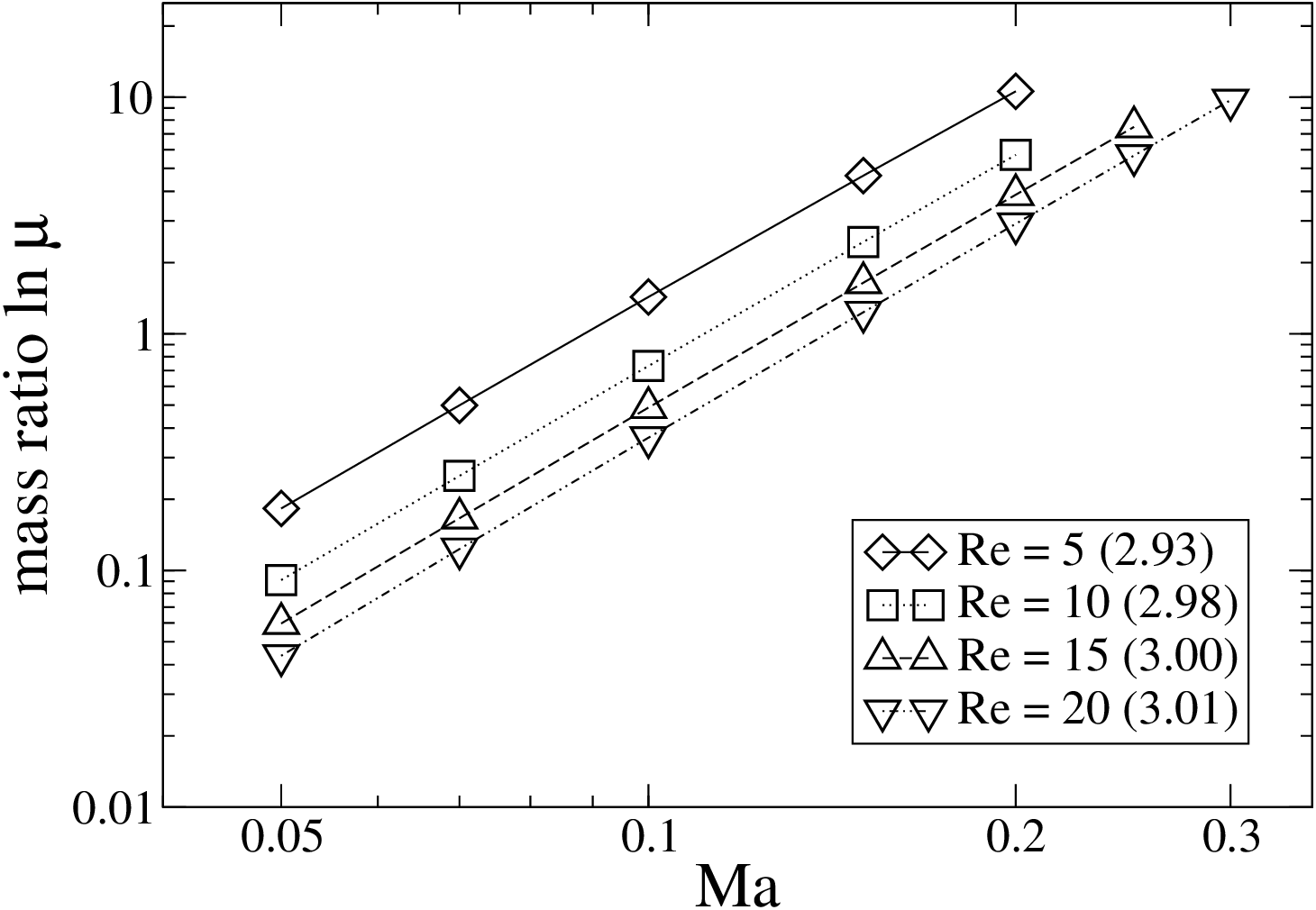}
\caption{Mean density $\bar \rho$ (top) in lattice units for $\text{Re} = 5$ and $\text{Ma} = 0.15$ as a function of time step: The simulation data (theoretical value $C_g \text{Ma}^3 / \text{Re} = 1.17 \cdot 10^{-4}$) is excellently reproduced by a simple exponential with $C_g \text{Ma}^3 / \text{Re} = 1.26 \cdot 10^{-4}$. Mass ratios $\ln \mu := \ln (M_{50000} / M_{10000})$ (bottom) for different $\text{Re}$ as function of the Mach number: The slopes are close to the theoretical value of $3$. Note that $\ln \mu$ itself is plotted logarithmically.}
\label{fig:massincrease}
\end{figure}

\subsection{Mass increase}
\label{subsec:results_massincrease}

A series of simulations with Reynolds numbers $5$, $10$, $15$ and $20$ and Mach numbers between $0.05$ and $0.3$ have been performed in a channel with dimensions $20 \times 22 \times 22$. In doing this we follow two points: (i) the verification of Eq.\ (\ref{eq:massfunction}) and (ii) the influence of the mass increase on the observables, especially the shear rate. The geometry factor $C_g$ for the simulations can be calculated using Eq.\ (\ref{eq:geometryfactor}). The number of time steps in all the simulations is $N_t = 5 \cdot 10^4$. Both the inlet / outlet boundary conditions proposed by Skordos and by Latt have been tested separately. Through the relaxation parameter $\tau$ the Mach number is adjusted.

The results for the time evolution of the mean density and the total mass increase after $5 \cdot 10^4$ time steps as a function of the Mach number are presented in figure \ref{fig:massincrease}. As a matter of fact, the time evolution of the mean density is well described by a simple exponential (top plot in figure \ref{fig:massincrease}). The recovered exponent for the presented example and the theoretical value only differ by $8\%$. An even better agreement can be found for the Mach number dependence (bottom plot in figure \ref{fig:massincrease}). We present a logarithmic plot of $\ln \mu$ versus the Mach number. $\mu > 0$ is defined as the ratio of the total mass at time step $5 \cdot 10^4$ and $1 \cdot 10^4$. Using the mass at $t = 1 \cdot 10^4$ instead of $t = 0$, transient effects are avoided. The predicted exponent of $3$ is nearly obtained in all simulations. As a result, the simple model presented in section \ref{sec:massincrease} is able to describe the mass increase during the simulation very well.

Some remarks should be made at this point. In figure \ref{fig:massincrease}, only the results using Latt's scheme are shown. The reason is that the results from Skordos' scheme basically are equivalent. We have observed, however, that the finite difference scheme by Skordos becomes unstable for values of $\tau$ larger than $\approx 1.8$. This is not very alarming, since usually LB simulations are performed at $\tau < 1$. The large values of $\tau$ have been necessary to increase the Mach number for fixed $\text{Re}$ and lattice size. Especially it has been shown that Skordos' scheme is superior to other boundary conditions when simulating flows with very large Reynolds numbers \cite{latt_straight_2008}. We would also like to underline that this analysis only makes sense as long as all parameters are in a range, where the LBM and the boundary conditions are reliable. Increasing $\tau$ even further leads to stronger deviations from the $\text{Ma}^3$-law (data not shown), but this behavior is of no practical interest.

The other important observation is that the relative errors Eq.\ (\ref{eq:l2error}) of the velocity and the shear rate do not depend on time, once steady state has been reached. Here, `steady state' is understood in the sense that, despite the increase of fluid density with time, the velocity field becomes time-independent. The possibility to define steady state alone shows that the mass increase does not affect the velocity. In fact, the errors of the velocity and the relevant shear component $S_{xy}$ remain constant within at least five significant digits. This observation even holds if the mass increase becomes huge, as for example in the case of $\text{Re} = 5$ and $\text{Ma} = 0.20$, where the mean density is $\bar \rho > 5.4 \cdot 10^4 $ after $5 \cdot 10^4$ time steps. The mass increase is, therefore, effectively harmless in the case of Poiseuille flow, which is a very encouraging result. It is an open question whether the mass increase affects the velocity and shear rate profiles in time-dependent simulations, since the distribution of the constantly instreaming mass may interfere with the time evolution of the physical system.


\section{Conclusions}
\label{sec:conclusions}

The shear stress --- beside the velocity --- has become an important observable of computer simulations, but an analysis of its convergence behavior has still been missing. We have shown that the shear stress obtained by the lattice Boltzmann method is second order accurate and thus behaves similar to the velocity. For a body force driven flow, a slope parameter of $-1.9$ has been found. Velocity boundary conditions, on the other hand, can spoil this convergence rate, but the effective scaling remains clearly better than first order ($-1.4$ and $-1.6$ for Skordos' and Latt's boundary conditions \cite{skordos_initial_1993, latt_straight_2008}, respectively). For this reason, it is not justified to denote the shear stress a first order quantity. It must be noted that the high symmetry of the Poiseuille flow may additionally decrease the error of the shear rate, but --- from a theoretical point of view --- the second order nature should not be affected by a modified geometry or time-dependent flows. This statement remains to be tested.

Like all numerical computations, lattice Boltzmann simulations require the setup of some basic dimensionless parameters, including the relaxation time $\tau$, the Mach number $\text{Ma}$ and the numerical grid size $N$. The choice of those parameters has a very significant impact on the accuracy of the simulations. Supporting Holdych et al.\ \cite{holdych_truncation_2004}, we come to the conclusion that it is in general recommended to use a relaxation parameter $\tau \approx 0.9$. This value leads to small velocity and shear errors. The accuracy of the simulations also depends on the Mach number, if velocity boundary conditions are used. Those boundary conditions become less reliable at large Mach numbers and may spoil especially the results for the shear stress. If large Mach numbers cannot be avoided, using a finite difference scheme for the shear rate is a good alternative, since it is more robust under those circumstances.

The use of velocity boundary conditions in the lattice Boltzmann method can lead to an artificial increase of the mass in the numerical grid. For three dimensional Poiseuille flow, we have demonstrated theoretically that the mass increase is only a function of the Mach number. If velocity boundary conditions at both inlet and outlet are used, the mass increase can only be reduced by choosing a smaller value for the Mach number. From a simple theoretical consideration it follows that the mass increase scales with $\text{Ma}^3$. This result has been confirmed with success in our simulations. The increase of the mass of the numerical grid does not influence the values of the velocity and the shear rate, since their definitions do not contain the local density. Although the mass increase is not physical, it basically poses no threat to the quality of the simulations. However, the velocity boundary conditions proposed by Skordos have been observed to become numerically unstable, if the relaxation parameter $\tau$ is larger than $\approx 1.8$.

\begin{acknowledgments}
This project has been supported by the DFG grant VA205/5-1.

Fruitful discussions with Jonas Latt and Orestis Malaspinas are gratefully acknowledged. \newline
\end{acknowledgments}



\begin{thebibliography}{10}%
\makeatletter
\providecommand \@ifxundefined [1]{%
 \ifx #1\undefined \expandafter \@firstoftwo
 \else \expandafter \@secondoftwo
\fi
}%
\providecommand \@ifnum [1]{%
 \ifnum #1\expandafter \@firstoftwo
 \else \expandafter \@secondoftwo
\fi
}%
\providecommand \enquote [1]{``#1''}%
\providecommand \bibnamefont  [1]{#1}%
\providecommand \bibfnamefont [1]{#1}%
\providecommand \citenamefont [1]{#1}%
\providecommand\href[0]{\@sanitize\@href}%
\providecommand\@href[1]{\endgroup\@@startlink{#1}\endgroup\@@href}%
\providecommand\@@href[1]{#1\@@endlink}%
\providecommand \@sanitize [0]{\begingroup\catcode`\&12\catcode`\#12\relax}%
\@ifxundefined \pdfoutput {\@firstoftwo}{%
 \@ifnum{\z@=\pdfoutput}{\@firstoftwo}{\@secondoftwo}%
}{%
 \providecommand\@@startlink[1]{\leavevmode}%
 \providecommand\@@endlink[0]{}%
}{%
 \providecommand\@@startlink[1]{%
  \leavevmode
  \pdfstartlink
   attr{/Border[0 0 1 ]/H/I/C[0 1 1]}%
   user{/Subtype/Link/A<</Type/Action/S/URI/URI(#1)>>}%
  \relax
 }%
 \providecommand\@@endlink[0]{\pdfendlink}%
}%
\providecommand \url  [0]{\begingroup\@sanitize \@url }%
\providecommand \@url [1]{\endgroup\@href {#1}{\urlprefix}}%
\providecommand \urlprefix [0]{URL }%
\providecommand \Eprint[0]{\href }%
\@ifxundefined \urlstyle {%
  \providecommand \doi [1]{doi:\discretionary{}{}{}#1}%
}{%
  \providecommand \doi [0]{doi:\discretionary{}{}{}\begingroup
  \urlstyle{rm}\Url }%
}%
\providecommand \doibase [0]{http://dx.doi.org/}%
\providecommand \Doi[1]{\href{\doibase#1}}%
\providecommand \bibAnnote [3]{%
  \BibitemShut{#1}%
  \begin{quotation}\noindent
    \textsc{Key:}\ #2\\\textsc{Annotation:}\ #3%
  \end{quotation}%
}%
\providecommand \bibAnnoteFile [2]{%
  \IfFileExists{#2}{\bibAnnote {#1} {#2} {\input{#2}}}{}%
}%
\providecommand \typeout [0]{\immediate \write \m@ne }%
\providecommand \selectlanguage [0]{\@gobble}%
\providecommand \bibinfo [0]{\@secondoftwo}%
\providecommand \bibfield [0]{\@secondoftwo}%
\providecommand \translation [1]{[#1]}%
\providecommand \BibitemOpen[0]{}%
\providecommand \bibitemStop [0]{}%
\providecommand \bibitemNoStop [0]{.\EOS\space}%
\providecommand \EOS [0]{\spacefactor3000\relax}%
\providecommand \BibitemShut [1]{\csname bibitem#1\endcsname}%
\bibitem{mcnamara_use_1988}%
  \BibitemOpen
  \bibfield{author}{%
  \bibinfo {author} {\bibfnamefont{G.~R.}\ \bibnamefont{McNamara}}\ and\
  \bibinfo {author} {\bibfnamefont{G.}~\bibnamefont{Zanetti}},\ }%
  \bibfield{journal}{%
  \bibinfo {journal} {Phys. Rev. Lett.}\ }%
  \textbf{\bibinfo {volume} {61}},\ \bibinfo {pages} {2332} (\bibinfo {year}
  {1988})%
  \bibAnnoteFile{NoStop}{mcnamara_use_1988}%
\bibitem{higuera_boltzmann_1989}%
  \BibitemOpen
  \bibfield{author}{%
  \bibinfo {author} {\bibfnamefont{F.~J.}\ \bibnamefont{Higuera}}\ and\
  \bibinfo {author} {\bibfnamefont{J.}~\bibnamefont{Jiménez}},\ }%
  \bibfield{journal}{%
  \bibinfo {journal} {Europhys. Lett.}\ }%
  \textbf{\bibinfo {volume} {9}},\ \bibinfo {pages} {663} (\bibinfo {year}
  {1989})%
  \bibAnnoteFile{NoStop}{higuera_boltzmann_1989}%
\bibitem{qian_lattice_1992}%
  \BibitemOpen
  \bibfield{author}{%
  \bibinfo {author} {\bibfnamefont{Y.~H.}\ \bibnamefont{Qian}}, \bibinfo
  {author} {\bibfnamefont{D.}~\bibnamefont{D'Humières}},\ and\ \bibinfo
  {author} {\bibfnamefont{P.}~\bibnamefont{Lallemand}},\ }%
  \bibfield{journal}{%
  \bibinfo {journal} {Europhys. Lett.}\ }%
  \textbf{\bibinfo {volume} {17}},\ \bibinfo {pages} {479} (\bibinfo {year}
  {1992})%
  \bibAnnoteFile{NoStop}{qian_lattice_1992}%
\bibitem{benzi_lattice_1992}%
  \BibitemOpen
  \bibfield{author}{%
  \bibinfo {author} {\bibfnamefont{R.}~\bibnamefont{Benzi}}, \bibinfo {author}
  {\bibfnamefont{S.}~\bibnamefont{Succi}},\ and\ \bibinfo {author}
  {\bibfnamefont{M.}~\bibnamefont{Vergassola}},\ }%
  \bibfield{journal}{%
  \bibinfo {journal} {Phys. Rep.}\ }%
  \textbf{\bibinfo {volume} {222}},\ \bibinfo {pages} {145} (\bibinfo {year}
  {1992})%
  \bibAnnoteFile{NoStop}{benzi_lattice_1992}%
\bibitem{ladd2001lbs}%
  \BibitemOpen
  \bibfield{author}{%
  \bibinfo {author} {\bibfnamefont{A.~J.~C.}\ \bibnamefont{Ladd}}\ and\
  \bibinfo {author} {\bibfnamefont{R.}~\bibnamefont{Verberg}},\ }%
  \bibfield{journal}{%
  \bibinfo {journal} {J. Stat. Phys.}\ }%
  \textbf{\bibinfo {volume} {104}},\ \bibinfo {pages} {1191} (\bibinfo {year}
  {2001})%
  \bibAnnoteFile{NoStop}{ladd2001lbs}%
\bibitem{chen_lattice_1998}%
  \BibitemOpen
  \bibfield{author}{%
  \bibinfo {author} {\bibfnamefont{S.}~\bibnamefont{Chen}}\ and\ \bibinfo
  {author} {\bibfnamefont{G.~D.}\ \bibnamefont{Doolen}},\ }%
  \bibfield{journal}{%
  \bibinfo {journal} {Annu. Rev. Fluid Mech.}\ }%
  \textbf{\bibinfo {volume} {30}},\ \bibinfo {pages} {329} (\bibinfo {year}
  {1998})%
  \bibAnnoteFile{NoStop}{chen_lattice_1998}%
\bibitem{kandhai_lattice-boltzmann_1998}%
  \BibitemOpen
  \bibfield{author}{%
  \bibinfo {author} {\bibfnamefont{D.}~\bibnamefont{Kandhai}}, \bibinfo
  {author} {\bibfnamefont{A.}~\bibnamefont{Koponen}}, \bibinfo {author}
  {\bibfnamefont{A.~G.}\ \bibnamefont{Hoekstra}}, \bibinfo {author}
  {\bibfnamefont{M.}~\bibnamefont{Kataja}}, \bibinfo {author}
  {\bibfnamefont{J.}~\bibnamefont{Timonen}},\ and\ \bibinfo {author}
  {\bibfnamefont{P.}~\bibnamefont{Sloot}},\ }%
  \bibfield{journal}{%
  \bibinfo {journal} {Comput. Phys. Commun.}\ }%
  \textbf{\bibinfo {volume} {111}},\ \bibinfo {pages} {14} (\bibinfo {year}
  {1998})%
  \bibAnnoteFile{NoStop}{kandhai_lattice-boltzmann_1998}%
\bibitem{succi_lattice_2001}%
  \BibitemOpen
  \bibfield{author}{%
  \bibinfo {author} {\bibfnamefont{S.}~\bibnamefont{Succi}},\ }%
  \emph{\bibinfo {title} {The Lattice Boltzmann Equation for Fluid Dynamics and
  Beyond}}\ (\bibinfo {publisher} {Oxford University Press},\ \bibinfo {year}
  {2001})\ ISBN \bibinfo {isbn} {978-0198503989},\ p.\ \bibinfo {pages} {368}%
  \bibAnnoteFile{NoStop}{succi_lattice_2001}%
\bibitem{sukop_lattice_2005}%
  \BibitemOpen
  \bibfield{author}{%
  \bibinfo {author} {\bibfnamefont{M.}~\bibnamefont{Sukop}}\ and\ \bibinfo
  {author} {\bibfnamefont{D.}~\bibnamefont{Thorne}},\ }%
  \emph{\bibinfo {title} {Lattice Boltzmann Modeling, an Introduction for
  Geoscientists and Engineers}}\ (\bibinfo {publisher} {Springer},\ \bibinfo
  {year} {2005})\ ISBN \bibinfo {isbn} {978-3540279815}%
  \bibAnnoteFile{NoStop}{sukop_lattice_2005}%
\bibitem{varnik_wetting_2008}%
  \BibitemOpen
  \bibfield{author}{%
  \bibinfo {author} {\bibfnamefont{F.}~\bibnamefont{Varnik}}, \bibinfo {author}
  {\bibfnamefont{P.}~\bibnamefont{Truman}}, \bibinfo {author}
  {\bibfnamefont{B.}~\bibnamefont{Wu}}, \bibinfo {author}
  {\bibfnamefont{P.}~\bibnamefont{Uhlmann}}, \bibinfo {author}
  {\bibfnamefont{D.}~\bibnamefont{Raabe}},\ and\ \bibinfo {author}
  {\bibfnamefont{M.}~\bibnamefont{Stamm}},\ }%
  \bibfield{journal}{%
  \bibinfo {journal} {Phys. Fluid}\ }%
  \textbf{\bibinfo {volume} {20}},\ \bibinfo {pages} {072104} (\bibinfo {month}
  {Jul.}\ \bibinfo {year} {2008})%
  \bibAnnoteFile{NoStop}{varnik_wetting_2008}%
\bibitem{varnik_roughness-induced_2007}%
  \BibitemOpen
  \bibfield{author}{%
  \bibinfo {author} {\bibfnamefont{F.}~\bibnamefont{Varnik}}, \bibinfo {author}
  {\bibfnamefont{D.}~\bibnamefont{Dorner}},\ and\ \bibinfo {author}
  {\bibfnamefont{D.}~\bibnamefont{Raabe}},\ }%
  \bibfield{journal}{%
  \bibinfo {journal} {J. Fluid Mech.}\ }%
  \textbf{\bibinfo {volume} {573}},\ \bibinfo {pages} {191} (\bibinfo {year}
  {2007})%
  \bibAnnoteFile{NoStop}{varnik_roughness-induced_2007}%
\bibitem{varnik_chaotic_2007}%
  \BibitemOpen
  \bibfield{author}{%
  \bibinfo {author} {\bibfnamefont{F.}~\bibnamefont{Varnik}}\ and\ \bibinfo
  {author} {\bibfnamefont{D.}~\bibnamefont{Raabe}},\ }%
  \bibfield{journal}{%
  \bibinfo {journal} {Mol. Simulat.}\ }%
  \textbf{\bibinfo {volume} {33}},\ \bibinfo {pages} {583} (\bibinfo {month}
  {Jun.}\ \bibinfo {year} {2007})%
  \bibAnnoteFile{NoStop}{varnik_chaotic_2007}%
\bibitem{varnik_scaling_2006}%
  \BibitemOpen
  \bibfield{author}{%
  \bibinfo {author} {\bibfnamefont{F.}~\bibnamefont{Varnik}}\ and\ \bibinfo
  {author} {\bibfnamefont{D.}~\bibnamefont{Raabe}},\ }%
  \bibfield{journal}{%
  \bibinfo {journal} {Modelling Simul. Mater. Sci. Eng.}\ }%
  \textbf{\bibinfo {volume} {14}},\ \bibinfo {pages} {857} (\bibinfo {year}
  {2006})%
  \bibAnnoteFile{NoStop}{varnik_scaling_2006}%
\bibitem{boyd_application_2005}%
  \BibitemOpen
  \bibfield{author}{%
  \bibinfo {author} {\bibfnamefont{J.}~\bibnamefont{Boyd}}, \bibinfo {author}
  {\bibfnamefont{J.}~\bibnamefont{Buick}}, \bibinfo {author}
  {\bibfnamefont{J.~A.}\ \bibnamefont{Cosgrove}},\ and\ \bibinfo {author}
  {\bibfnamefont{P.}~\bibnamefont{Stansell}},\ }%
  \bibfield{journal}{%
  \bibinfo {journal} {Phys. Med. Biol.}\ }%
  \textbf{\bibinfo {volume} {50}},\ \bibinfo {pages} {4783} (\bibinfo {year}
  {2005})%
  \bibAnnoteFile{NoStop}{boyd_application_2005}%
\bibitem{xiu-ying_three-dimensional_2008}%
  \BibitemOpen
  \bibfield{author}{%
  \bibinfo {author} {\bibfnamefont{X.-Y.}\ \bibnamefont{Kang}}, \bibinfo
  {author} {\bibfnamefont{Y.-P.}\ \bibnamefont{Ji}}, \bibinfo {author}
  {\bibfnamefont{D.-H.}\ \bibnamefont{Liu}},\ and\ \bibinfo {author}
  {\bibfnamefont{Y.-J.}\ \bibnamefont{Jin}},\ }%
  \bibfield{journal}{%
  \bibinfo {journal} {Chinese Phys. B}\ }%
  \textbf{\bibinfo {volume} {17}},\ \bibinfo {pages} {1041} (\bibinfo {year}
  {2008})%
  \bibAnnoteFile{NoStop}{xiu-ying_three-dimensional_2008}%
\bibitem{schneider_shear-induced_2007}%
  \BibitemOpen
  \bibfield{author}{%
  \bibinfo {author} {\bibfnamefont{S.~W.}\ \bibnamefont{Schneider}}, \bibinfo
  {author} {\bibfnamefont{S.}~\bibnamefont{Nuschele}}, \bibinfo {author}
  {\bibfnamefont{A.}~\bibnamefont{Wixforth}}, \bibinfo {author}
  {\bibfnamefont{C.}~\bibnamefont{Gorzelanny}}, \bibinfo {author}
  {\bibfnamefont{A.}~\bibnamefont{Alexander-Katz}}, \bibinfo {author}
  {\bibfnamefont{R.~R.}\ \bibnamefont{Netz}},\ and\ \bibinfo {author}
  {\bibfnamefont{M.~F.}\ \bibnamefont{Schneider}},\ }%
  \bibfield{journal}{%
  \bibinfo {journal} {Proc. Natl. Acad. Sci. U.S.A.}\ }%
  \textbf{\bibinfo {volume} {104}},\ \bibinfo {pages} {7899} (\bibinfo {year}
  {2007})%
  \bibAnnoteFile{NoStop}{schneider_shear-induced_2007}%
\bibitem{frisch_lattice-gas_1986}%
  \BibitemOpen
  \bibfield{author}{%
  \bibinfo {author} {\bibfnamefont{U.}~\bibnamefont{Frisch}}, \bibinfo {author}
  {\bibfnamefont{B.}~\bibnamefont{Hasslacher}},\ and\ \bibinfo {author}
  {\bibfnamefont{Y.}~\bibnamefont{Pomeau}},\ }%
  \bibfield{journal}{%
  \bibinfo {journal} {Phys. Rev. Lett.}\ }%
  \textbf{\bibinfo {volume} {56}},\ \bibinfo {pages} {1505} (\bibinfo {year}
  {1986})%
  \bibAnnoteFile{NoStop}{frisch_lattice-gas_1986}%
\bibitem{hou_simulation_1995}%
  \BibitemOpen
  \bibfield{author}{%
  \bibinfo {author} {\bibfnamefont{S.}~\bibnamefont{Hou}}, \bibinfo {author}
  {\bibfnamefont{Q.}~\bibnamefont{Zou}}, \bibinfo {author}
  {\bibfnamefont{S.}~\bibnamefont{Chen}}, \bibinfo {author}
  {\bibfnamefont{G.}~\bibnamefont{Doolen}},\ and\ \bibinfo {author}
  {\bibfnamefont{A.~C.}\ \bibnamefont{Cogley}},\ }%
  \bibfield{journal}{%
  \bibinfo {journal} {J. Comput. Phys.}\ }%
  \textbf{\bibinfo {volume} {118}},\ \bibinfo {pages} {329} (\bibinfo {year}
  {1995})%
  \bibAnnoteFile{NoStop}{hou_simulation_1995}%
\bibitem{qian_higher-order_2000}%
  \BibitemOpen
  \bibfield{author}{%
  \bibinfo {author} {\bibfnamefont{Y.-H.}\ \bibnamefont{Qian}}\ and\ \bibinfo
  {author} {\bibfnamefont{Y.}~\bibnamefont{Zhou}},\ }%
  \bibfield{journal}{%
  \bibinfo {journal} {Phys. Rev. E}\ }%
  \textbf{\bibinfo {volume} {61}},\ \bibinfo {pages} {2103} (\bibinfo {year}
  {2000})%
  \bibAnnoteFile{NoStop}{qian_higher-order_2000}%
\bibitem{chikatamarla_entropy_2006}%
  \BibitemOpen
  \bibfield{author}{%
  \bibinfo {author} {\bibfnamefont{S.~S.}\ \bibnamefont{Chikatamarla}}\ and\
  \bibinfo {author} {\bibfnamefont{I.~V.}\ \bibnamefont{Karlin}},\ }%
  \bibfield{journal}{%
  \bibinfo {journal} {Phys. Rev. Lett.}\ }%
  \textbf{\bibinfo {volume} {97}},\ \bibinfo {pages} {190601} (\bibinfo {year}
  {2006})%
  \bibAnnoteFile{NoStop}{chikatamarla_entropy_2006}%
\bibitem{siebert_lattice_2008}%
  \BibitemOpen
  \bibfield{author}{%
  \bibinfo {author} {\bibfnamefont{D.~N.}\ \bibnamefont{Siebert}}, \bibinfo
  {author} {\bibfnamefont{L.~A.~H.}\ \bibnamefont{Jr.}},\ and\ \bibinfo
  {author} {\bibfnamefont{P.~C.}\ \bibnamefont{Philippi}},\ }%
  \bibfield{journal}{%
  \bibinfo {journal} {Phys. Rev. E}\ }%
  \textbf{\bibinfo {volume} {77}},\ \bibinfo {pages} {026707} (\bibinfo {year}
  {2008})%
  \bibAnnoteFile{NoStop}{siebert_lattice_2008}%
\bibitem{nie_galilean_2008}%
  \BibitemOpen
  \bibfield{author}{%
  \bibinfo {author} {\bibfnamefont{X.~B.}\ \bibnamefont{Nie}}, \bibinfo
  {author} {\bibfnamefont{X.}~\bibnamefont{Shan}},\ and\ \bibinfo {author}
  {\bibfnamefont{H.}~\bibnamefont{Chen}},\ }%
  \bibfield{journal}{%
  \bibinfo {journal} {Europhys. Lett.}\ }%
  \textbf{\bibinfo {volume} {81}},\ \bibinfo {pages} {34005} (\bibinfo {year}
  {2008})%
  \bibAnnoteFile{NoStop}{nie_galilean_2008}%
\bibitem{mei_force_2002}%
  \BibitemOpen
  \bibfield{author}{%
  \bibinfo {author} {\bibfnamefont{R.}~\bibnamefont{Mei}}, \bibinfo {author}
  {\bibfnamefont{D.}~\bibnamefont{Yu}}, \bibinfo {author}
  {\bibfnamefont{W.}~\bibnamefont{Shyy}},\ and\ \bibinfo {author}
  {\bibfnamefont{L.-S.}\ \bibnamefont{Luo}},\ }%
  \bibfield{journal}{%
  \bibinfo {journal} {Phys. Rev. E}\ }%
  \textbf{\bibinfo {volume} {65}},\ \bibinfo {pages} {041203} (\bibinfo {year}
  {2002})%
  \bibAnnoteFile{NoStop}{mei_force_2002}%
\bibitem{holdych_truncation_2004}%
  \BibitemOpen
  \bibfield{author}{%
  \bibinfo {author} {\bibfnamefont{D.~J.}\ \bibnamefont{Holdych}}, \bibinfo
  {author} {\bibfnamefont{D.~R.}\ \bibnamefont{Noble}}, \bibinfo {author}
  {\bibfnamefont{J.~G.}\ \bibnamefont{Georgiadis}},\ and\ \bibinfo {author}
  {\bibfnamefont{R.~O.}\ \bibnamefont{Buckius}},\ }%
  \bibfield{journal}{%
  \bibinfo {journal} {J. Comput. Phys.}\ }%
  \textbf{\bibinfo {volume} {193}},\ \bibinfo {pages} {595} (\bibinfo {year}
  {2004})%
  \bibAnnoteFile{NoStop}{holdych_truncation_2004}%
\bibitem{haberman_applied_2004}%
  \BibitemOpen
  \bibfield{author}{%
  \bibinfo {author} {\bibfnamefont{R.}~\bibnamefont{Haberman}},\ }%
  \emph{\bibinfo {title} {Applied Partial Differential Equations: with Fourier
  Series and Boundary Value Problems}}\ (\bibinfo {publisher} {Pearson Prentice
  Hall},\ \bibinfo {year} {2004})\ ISBN \bibinfo {isbn} {978-0130652430},\ p.\
  \bibinfo {pages} {769}%
  \bibAnnoteFile{NoStop}{haberman_applied_2004}%
\bibitem{he_analytic_1997}%
  \BibitemOpen
  \bibfield{author}{%
  \bibinfo {author} {\bibfnamefont{X.}~\bibnamefont{He}}, \bibinfo {author}
  {\bibfnamefont{Q.}~\bibnamefont{Zou}}, \bibinfo {author}
  {\bibfnamefont{L.-S.}\ \bibnamefont{Luo}},\ and\ \bibinfo {author}
  {\bibfnamefont{M.}~\bibnamefont{Dembo}},\ }%
  \bibfield{journal}{%
  \bibinfo {journal} {J. Stat. Phys.}\ }%
  \textbf{\bibinfo {volume} {87}},\ \bibinfo {pages} {115} (\bibinfo {year}
  {1997})%
  \bibAnnoteFile{NoStop}{he_analytic_1997}%
\bibitem{guo_discrete_2002}%
  \BibitemOpen
  \bibfield{author}{%
  \bibinfo {author} {\bibfnamefont{Z.}~\bibnamefont{Guo}}, \bibinfo {author}
  {\bibfnamefont{C.}~\bibnamefont{Zheng}},\ and\ \bibinfo {author}
  {\bibfnamefont{B.}~\bibnamefont{Shi}},\ }%
  \bibfield{journal}{%
  \bibinfo {journal} {Phys. Rev. E}\ }%
  \textbf{\bibinfo {volume} {65}},\ \bibinfo {pages} {046308} (\bibinfo {year}
  {2002})%
  \bibAnnoteFile{NoStop}{guo_discrete_2002}%
\bibitem{skordos_initial_1993}%
  \BibitemOpen
  \bibfield{author}{%
  \bibinfo {author} {\bibfnamefont{P.~A.}\ \bibnamefont{Skordos}},\ }%
  \bibfield{journal}{%
  \bibinfo {journal} {Phys. Rev. E}\ }%
  \textbf{\bibinfo {volume} {48}},\ \bibinfo {pages} {4823} (\bibinfo {year}
  {1993})%
  \bibAnnoteFile{NoStop}{skordos_initial_1993}%
\bibitem{latt_straight_2008}%
  \BibitemOpen
  \bibfield{author}{%
  \bibinfo {author} {\bibfnamefont{J.}~\bibnamefont{Latt}}, \bibinfo {author}
  {\bibfnamefont{B.}~\bibnamefont{Chopard}}, \bibinfo {author}
  {\bibfnamefont{O.}~\bibnamefont{Malaspinas}}, \bibinfo {author}
  {\bibfnamefont{M.}~\bibnamefont{Deville}},\ and\ \bibinfo {author}
  {\bibfnamefont{A.}~\bibnamefont{Michler}},\ }%
  \bibfield{journal}{%
  \bibinfo {journal} {Phys. Rev. E}\ }%
  \textbf{\bibinfo {volume} {77}},\ \bibinfo {pages} {056703} (\bibinfo {year}
  {2008})%
  \bibAnnoteFile{NoStop}{latt_straight_2008}%
\bibitem{reider_accuracy_1995}%
  \BibitemOpen
  \bibfield{author}{%
  \bibinfo {author} {\bibfnamefont{M.~B.}\ \bibnamefont{Reider}}\ and\ \bibinfo
  {author} {\bibfnamefont{J.~D.}\ \bibnamefont{Sterling}},\ }%
  \bibfield{journal}{%
  \bibinfo {journal} {Comput. Fluids}\ }%
  \textbf{\bibinfo {volume} {24}},\ \bibinfo {pages} {459} (\bibinfo {year}
  {1995})%
  \bibAnnoteFile{NoStop}{reider_accuracy_1995}%
\bibitem{cates_physical_2005}%
  \BibitemOpen
  \bibfield{author}{%
  \bibinfo {author} {\bibfnamefont{M.~E.}\ \bibnamefont{Cates}}, \bibinfo
  {author} {\bibfnamefont{J.-C.}\ \bibnamefont{Desplat}}, \bibinfo {author}
  {\bibfnamefont{P.}~\bibnamefont{Stansell}},\ and\ \bibinfo {author}
  {\bibnamefont{et~al.}},\ }%
  \bibfield{journal}{%
  \bibinfo {journal} {Phil. Trans. R. Soc. A}\ }%
  \textbf{\bibinfo {volume} {363}},\ \bibinfo {pages} {1917} (\bibinfo {year}
  {2005})%
  \bibAnnoteFile{NoStop}{cates_physical_2005}%
\bibitem{chopard_mass_2003}%
  \BibitemOpen
  \bibfield{author}{%
  \bibinfo {author} {\bibfnamefont{B.}~\bibnamefont{Chopard}}\ and\ \bibinfo
  {author} {\bibfnamefont{A.}~\bibnamefont{Dupuis}},\ }%
  \bibfield{journal}{%
  \bibinfo {journal} {Int. J. Mod. Phys. B}\ }%
  \textbf{\bibinfo {volume} {17}},\ \bibinfo {pages} {103} (\bibinfo {year}
  {2003})%
  \bibAnnoteFile{NoStop}{chopard_mass_2003}%
\bibitem{hollis_enhanced_2006}%
  \BibitemOpen
  \bibfield{author}{%
  \bibinfo {author} {\bibfnamefont{A.}~\bibnamefont{Hollis}}, \bibinfo {author}
  {\bibfnamefont{I.}~\bibnamefont{Halliday}},\ and\ \bibinfo {author}
  {\bibfnamefont{C.~M.}\ \bibnamefont{Care}},\ }%
  \bibfield{journal}{%
  \bibinfo {journal} {J. Phys. A: Math. Gen.}\ }%
  \textbf{\bibinfo {volume} {39}},\ \bibinfo {pages} {10589} (\bibinfo {year}
  {2006})%
  \bibAnnoteFile{NoStop}{hollis_enhanced_2006}%
\bibitem{he_lattice_1997}%
  \BibitemOpen
  \bibfield{author}{%
  \bibinfo {author} {\bibfnamefont{X.}~\bibnamefont{He}}\ and\ \bibinfo
  {author} {\bibfnamefont{L.-S.}\ \bibnamefont{Luo}},\ }%
  \bibfield{journal}{%
  \bibinfo {journal} {J. Stat. Phys.}\ }%
  \textbf{\bibinfo {volume} {88}},\ \bibinfo {pages} {927} (\bibinfo {year}
  {1997})%
  \bibAnnoteFile{NoStop}{he_lattice_1997}%
\bibitem{guo_lattice_2000}%
  \BibitemOpen
  \bibfield{author}{%
  \bibinfo {author} {\bibfnamefont{Z.}~\bibnamefont{Guo}}, \bibinfo {author}
  {\bibfnamefont{B.}~\bibnamefont{Shi}},\ and\ \bibinfo {author}
  {\bibfnamefont{N.}~\bibnamefont{Wang}},\ }%
  \bibfield{journal}{%
  \bibinfo {journal} {J. Comput. Phys.}\ }%
  \textbf{\bibinfo {volume} {165}},\ \bibinfo {pages} {288} (\bibinfo {year}
  {2000})%
  \bibAnnoteFile{NoStop}{guo_lattice_2000}%
\end{thebibliography}

%

\end{document}